 \documentclass[twocolumn]{emulateapj}

\shorttitle{8--13 $\mu$m spectroscopy of YSOs}
\shortauthors{Kessler-Silacci et al.}

\begin{document}


\title{8--13 $\mu$m spectroscopy of YSOs: Evolution of the silicate feature}


\author{Jacqueline E. Kessler-Silacci\altaffilmark{1,2}, Lynne A. Hillenbrand\altaffilmark{3}, Geoffrey A. Blake\altaffilmark{2,4},\\
 and Michael R. Meyer\altaffilmark{5}}






\altaffiltext{1}{Department of Astronomy, The University of Texas at Austin C1400, Austin, TX, 78712, USA.} 
\altaffiltext{2}{Division of Chemistry and Chemical Engineering, California Institute of Technology 105-24, Pasadena, CA 91125, USA.}
\altaffiltext{3}{Division of Physics, Math and Astronomy, California Institute of Technology 105-24, Pasadena, CA 91125, USA.} 
\altaffiltext{4}{Division of  Geological and Planetary Sciences, California Institute of Technology 150-21, Pasadena, CA 
	91125, USA.}
\altaffiltext{5}{Steward Observatory, The University of Arizona, Tucson, AZ 85721, USA.}


\begin{abstract} 
Silicate features arising from material around pre-main sequence stars are useful probes of the star and planet formation process.
In order to investigate possible connections between dust processing and disk properties, 
8--13 $\mu$m spectra of 34 young stars, exhibiting a range of circumstellar environments and including spectral types A to M, were obtained using the Long Wavelength Spectrometer at the W.  M. Keck Observatory.  The broad 9.7 $\mu$m amorphous silicate (Si-O stretching) feature which dominates this wavelength regime evolves from absorption in young, embedded sources, to emission in optically revealed stars, and to complete absence in older ``debris'' disk systems for both low- and intermediate-mass stars.  This is similar to the evolutionary pattern seen in ISO observations of high/intermediate-mass YSOs.  The peak wavelength and FWHM are centered about 9.7 $\mu$m and $\sim$2.3 $\mu$m, respectively, corresponding to amorphous olivine, with a larger spread in FWHM for embedded sources and in peak wavelength for disks. In a few of our objects that have been previously identified as class I low-mass YSOs, the observed silicate feature is more complex, with absorption near 9.5 $\mu$m and emission peaking around 10 $\mu$m. 
Although most of the emission spectra show broad classical features attributed to amorphous silicates, small variations in the shape/strength may be linked to dust processing, including grain growth and/or silicate crystallization. 
 For some of the Herbig Ae stars in the sample, the broad emission feature has an additional bump near 11.3 $\mu$m, similar to the emission from crystalline forsterite seen in comets and the debris disk $\beta$ Pictoris.  Only one of the low-mass stars, Hen 3-600A, and one Herbig Ae star, HD 179218 clearly show strong, narrow emission near 11.3 $\mu$m.  We study quantitatively the evidence for evolutionary trends in the 8--13 $\mu$m spectra through a variety of spectral shape diagnostics.  Based on the lack of correlation between these diagnostics and broad-band infrared luminosity characteristics for silicate emission sources we conclude that although spectral signatures of dust processing are present, they can not be connected clearly to disk evolutionary stage (for optically thick disks) or optical depth (for optically thin disks).  The diagnostics of silicate absorption features (other than the central wavelength of the feature), however, are tightly correlated with optical depth and thus do not probe silicate grain properties.
\end{abstract}




\section{INTRODUCTION: SILICATES AND STAR FORMATION}

Circumstellar disks are found around most, if not all, young
stars.   At least 5 $\%$ of these 
circumstellar disks are likely planet-forming
\citep{marcy_pp4} and it is commonly believed that our
solar system formed from a solar nebula similar to circumstellar
disks found around young stars (Russell et al., Met. and Early
Solar System II in press).  Initially these disks are composed of
gas and dust from the interstellar medium.  Theoretical
models predict that collisions between grains and subsequent
coagulation result in modified grain size distributions and
eventually the growth of planetesimals \citep{weidenschilling_pp3}.  
However, observational evidence of this process
remains elusive \citep{beckwith_pp4}.  One way to gain insight
into the process of grain growth, planetesimal evolution, and
eventually planet formation, is to constrain the effects of these
processes on the observed mid-infrared spectra from circumstellar
dust.  Grain growth and changes in composition due to dust processing
in circumstellar disks can have dramatic effects on the observed
spectrum of the dust in emission and absorption.  In this contribution,
we attempt to quantify those effects and place them in the context
of modern theories of disk evolution and planet formation.

Dust in the interstellar medium, and thus in the early stages of star formation, consists primarily of oxygen-rich silicates and carbon-rich grains \citep[for summary, see][]{pollack_94} from a wide range of sources, including stellar atmospheres, novae, supernovae. These silicates are either olivines (Mg$_{2x}$Fe$_{(1-x)}$SiO$_4$), ranging from fayalite ($x=0$) to forsterite ($x=1$), or pyroxenes (Mg$_{x}$Fe$_{(1-x)}$SiO$_3$), ranging from ferrosillite ($x=0$) to enstatite ($x=1$).  Silicate condensates formed in supernovae and during star formation are primarily Fe-rich and amorphous, but at high temperatures, Mg-rich crystalline silicates (i.e., pure forsterite and enstatite) are produced. The fraction of pyroxenes versus olivines depends on both temperature and pressure.  For a given pressure, the pyroxenes (e.g., enstatite) dominate at low temperatures and becomes less abundant with increasing temperature (at equilibrium).  The olivines (e.g., forsterite) form a significant fraction of the mixture only for a narrow region in pressure-temperature space. 

The low temperatures of the ISM are expected to result in amorphous mixtures of olivines and pyroxenes with olivine the dominant component.
Indeed, strong
smooth features near 9.7 (Si-O stretching mode) and 18.5 $\mu$m (O-Si-O bending mode) are observed in absorption along lines of sight through the diffuse interstellar medium \citep{whittet_97, bouwman_01}.
Molecular clouds also possess a  mixture of olivines and pyroxenes, 
which are expected to remain amorphous due to cosmic ray impacts; a broad structureless band centered near 10 $\mu$m is observed toward several objects \citep{wilner_82, pegourie_85, whittet_88, bowey_98}.  
Observations by \citet{bowey_02} suggest a shift from amorphous olivine dominated grains in diffuse-medium environments to amorphous pyroxenes in molecular clouds and YSOs, indicating that some dust processing is occurring during the early stages of star formation. Further processing is evidenced by the presence of crystalline silicates in the dust surrounding some main sequence stars, with amorphous olivine and forsterite as the dominant components \citep[see e.g.,][]{knacke_f93}.  
Comets also contain both amorphous and crystalline silicates \citep[see review by][]{wooden_02}, with crystalline fractions that are too high (up to 30$\%$ by mass) to be accounted for by the ISM \citep[where the crystalline fraction is $<$~2$\%$;][]{kemper_04}, indicating that crystallization occurred in the proto-solar nebula. 

Differences in the silicate spectra of interstellar versus solar-system dust 
suggest an evolutionary sequence in which the grain composition 
changes from amorphous olivine  to crystalline forsterite 
during the star and solar system formation 
process. Dust grains in accretion disks around young stars
therefore must be significantly modified with respect to those of the ISM and 
are thus excellent laboratories 
for the study of {\it in situ} crystallization of silicates.  
If crystallization proceeds via annealing, then 
silicates in the outer disk are expected to accrete icy mantles and retain cores that are similar to ISM dust (i.e., amorphous olivine and pyroxene), while in the inner disk (T $\gtrsim$ 800 K),  microcrystalline lattice structures on 0.001 $\mu$m size scales  can form within silicate grains \citep{gail_98}. If annealing continues for a sufficiently long time, or occurs in a sufficiently hot environment, crystallization spreads from these critical nuclei over the entire grain. At disk radii corresponding to temperatures above $\sim$ 1000 K, 0.1--1 $\mu$m grains should be completely crystallized \citep{gail_98}  and larger dust grains should possess regions of localized crystallization detectable in IR spectra \citep{jager_94}.  Increases in temperature are also predicted to result in conversion from pyroxenes to olivines. Thus we would expect both crystallization and composition (i.e., the content of Fe vs. Mg and olivine vs. pyroxene) of the silicates in disks to be dependent on exposure to stellar radiation (e.g., location in the disk, disk geometry, stellar type) and/or the duration of processing (e.g., amount of disk turbulence, disk age).  

High spectral resolution studies of the warm dust emission near 10 $\mu$m
are vital in understanding the composition and processing of dust 
in the planet-forming region (1--10 AU) of circumstellar disks around 
young stars. The dominant feature in the 8--13~$\mu$m spectral region
accessible from the ground is from amorphous 
olivine (peaking at 9.5 $\mu$m), with weaker features arising from 
silica (SiO$_2$; 8.6 $\mu$m) and crystalline silicates including 
enstatite (MgSiO$_3$; 9.2, 10.4 $\mu$m) and forsterite (Mg$_2$SiO$_4$; 10.2, 
11.3 $\mu$m)\footnotemark.
\footnotetext{The peak wavelengths described above are for 0.1 $\mu$m grains \citep{meeus_01}.}
The transition from amorphous olivine to crystalline forsterite 
results in a predicted shift of the silicate emission feature from 9.7 to 11.3 $\mu$m. However,
analysis of the 10 $\mu$m silicate feature is complicated by the fact that the 
strength and shape of the emission can not be uniquely correlated with 
abundance or composition. The presence of organics on 
grain surfaces results in increased opacity at $\lambda$ $<$ 6 $\mu$m 
and decreased
contrast between the silicate band at 10~$\mu$m and the adjacent continuum 
\citep{dalessio_c01}. Grain size has a similar effect and for large 
grains ($a_{max}>10$~$\mu$m) the silicate bands disappear almost entirely 
\citep{wolf_03}. Additionally, \citet{fabian_01} suggest 
that grain shape also needs to be taken into account in the interpretation of 
silicate spectra, particularly for strong crystalline bands, for which 
a combination of spherical and ellipsoidal grains may be required.  
Disk morphology and optical depth also influence the shape/strength 
of the silicate feature \citep{meeus_01, bouwman_03}.  Despite 
difficulties in constructing unique interpretations, 
a number of ground- and space-based studies of the
disks around T Tauri and Herbig Ae/Be (hereafter referred to as HAEBE) stars
have been conducted, yielding important constraints on the evolution of circumstellar material.

Early studies of young low-mass stars \citep{cohen_80, cohen_85, wooden_94, hanner_b95, hanner_b98} show a single, rather sharp peak between 9 and 10 $\mu$m, resembling molecular clouds and the diffuse ISM. 
ISO spectra of  young low-mass stars in 
Chameleon 
\citep{natta_00} display 10 $\mu$m silicate emission that is significantly broader and stronger 
than that from the diffuse ISM, but consistent with the emission 
predicted by a simple radiation transfer model of $<$1 $\mu$m amorphous 
pyroxene and/or olivine grains in a flared disk.
More recent surveys \citep[e.g.,][]{sitko_99, sitko_00, sylvester_m00, 
honda_03, meeus_03, przygodda_03, vanboekel_03}, however, have found 
evidence in a few low-mass sources for a more trapezoidal shaped 8--13 $\mu$m amorphous silicate feature  or even a pronounced 11.3 $\mu$m crystalline 
silicate (forsterite) bump, indicating that at least some dust processing has 
taken place in these objects. 

HAEBE stars were studied more extensively with ISO and,
based on 2--45 $\mu$m ISO spectra of HAEBE stars spanning a range of ages,
an evolutionary sequence has been suggested \citep{meeus_99}. 
In this sequence, rising featureless dense cloud core spectra proceed
to strong amorphous 10 $\mu$m silicate absorption from embedded protostars.
These are followed by optically thick protoplanetary disks with spectra 
showing warm 10 $\mu$m silicate dust emission and prominent cold dust 
emission longward of 15 $\mu$m, then spectra with 
only warm dust interpreted as 
arising from disks undergoing planet formation, and finally the most
evolved sources with nearly photospheric spectra,
indicating disk dispersal.
In addition to this general evolutionary trend,  
ISO detected crystalline silicate emission 
bands (11.3, 27.5, 33.5, 35.8 and 70 $\mu$m)  toward a modest sample of 
HAEBE star+disk systems \citep{molster_99, meeus_01}. 
The longer wavelength emission bands are highly 
dependent on the structure of the dust. These studies show 
large variations in dust crystallinity and composition among YSOs of similar 
age and stellar properties \citep{vandenancker_00, meeus_01}, indicating the 
necessity of large samples to test scenarios of disk 
evolution.  

Objects that may be in transition between primordial dust and debris dust 
(generated through collisions of planetesimals), or which are fully 
debris, have also been investigated \citep{sylvester_96, 
sylvester_m00} revealing generally photospheric or continuum excess spectra. 
The 10 $\mu$m spectrum of the 12$^{+8}_{-4}$ Myr old \citep{zuckerman_01} 
A star $\beta$ Pictoris, however,
 exhibits a remarkable resemblance to that of solar system 
grains and comets, suggesting a mixture of crystalline and amorphous material
\citep{knacke_f93, reach_03}.
Direct imaging of $\beta$ Pictoris shows a warped disk that
is potentially caused by collisions of planetesimals and may be
shaped by recent planet formation \citep{kalas_95, mouillet_97, augereau_01}.  
Spatially resolved 10 $\mu$m spectroscopy of $\beta$ Pictoris 
\citep{weinberger_03} revealed that the silicate emission arises only 
from dust within 20 AU of the star, compared to the 77 AU emitting region
in the continuum.  Furthermore, the 
mixture of amorphous and crystalline dust appears to remain constant, 
suggesting that the crystallization process may be 
acting uniformly as a function of radius within the planet formation region.

Observations of silicate emission toward young stellar objects in a variety
of evolutionary stages can provide valuable information about the modification 
of dust during solar system formation.  In the present study, 8--13 $\mu$m 
spectra were obtained for 34 low- and 
intermediate-mass young stars, ranging from embedded protostars to debris disk 
systems. Additional photometry at 10.7~$\mu$m was obtained for a subset of the 
sample in order 
to facilitate flux calibration of the spectra.  In $\S$2, the sample selection 
criteria are discussed.  The spectroscopic and photometric observations and 
data reduction are described in $\S$3.  SEDs compiled from the literature 
are presented in $\S$4, along with implications 
regarding the evolutionary categories within the sample.  In $\S$5, the 
8--13~$\mu$m spectra are discussed qualitatively and in $\S$6 a quantitative 
analysis of the silicate emission spectra is presented. Conclusions and 
suggestions of future work are discussed in $\S$7.  The appendices contain the 
references for the SEDs (A) and a short discussion of variability in the 
silicate feature (B).

\section{SAMPLE SELECTION}

\begin{deluxetable*}{lllcccccc}
\tablewidth{0pt}
\tablecaption{Observations \label{tab:obs}}
\tablehead{\colhead{} & \colhead{RA} & \colhead{DEC} & \colhead{Observation} & \colhead{Reference} & \colhead{Aperture}  & \colhead{} \\
\colhead{Source}           & \colhead{(J2000)} & \colhead{(J2000)} & \colhead{Date} 	& \colhead{Star}         & \colhead{(pixels)}  & \colhead{{LSEC}?}} 
\startdata
49 Cet           			&01 34 37.78   &-15 40 34.9     & 2000 Nov 8   			& $\alpha$ Tau          & 6       & n   \\   *
HD 17925         			&02 52 32.129  &-12 46 10.972   & 2000 Feb 21			& HR 1017               & 6       & y   \\   *
HD 22049         			&03 32 55.844  &-09 27 29.744   & 2000 Dec 10			& HR 1017               & 6       & y   \\   *
IRAS 04016+2610 			&  04 04 42.85 &  +26 18 56.3	& 2000 Dec 9 			& HR 617            	& 6       & y   \\   *
IRAS 04108+2803B \tablenotemark{a} 	&  04 13 52.9  &  +28 11 23	& 2000 Dec 10			& HR 1017  	    	& 6       & y   \\   *
IRAS 04169+2702	 			&  04 19 59.24 &  +27 09 58.6   & 2000 Dec 10			& HR 1017           	& 6       & y   \\   *
IRAS 04181+2654A/B \tablenotemark{a} 	&  04 21 11.42 &  +27 01 08.9   & 2000 Dec 10			& HR 531   	    	& 6       & y   \\   *
IRAS 04239+2436  			&  04 26 57.1  &  +24 43 36     & 2000 Dec 10			& HR 1017           	& 6       & y   \\   *
IRAS 04248+2612	 			&  04 27 56.7  &  +26 19 20     & 2000 Dec  9			& HR 617            	& 6       & y   \\   *
IRAS 04264+2433 \tablenotemark{a}	&  04 29 07.68 &  +24 43 50.1   & 2000 Dec  9			& HR 1017           	& 6       & y   \\   *
Haro 6-10A	 			&  04 29 24.39 &  +24 33 02.1	& 2000 Dec 10			& HR 1017           	& 6       & y   \\   *
Haro 6-10B	 			&  04 29 24.39 &  +24 33 02.1	& 2000 Dec 10			& HR 1017	    	& 6       & y   \\   *
IRAS 04287+1801	 			&  04 31 33.6  &  +18 08 15     & 2000 Dec  9			& HR 1708           	& 6       & y   \\   *
IRAS 04295+2251 			&  04 32 32.07 &  +22 57 30.3   & 2000 Dec  9			& HR 617                & 6       & y   \\   *
AA Tau           			&04 34 55.2 	 &+24 28 52	& 2000 Feb 21			& HR 2990               & 6       & y   \\   *
IRAS 04325+2402	\tablenotemark{b}	&  04 35 33.0    &  +24 08 14	& 2000 Dec  9			& ...			& ...	  & ... \\   *
LkCa 15 \tablenotemark{a}          	&04 39 17.8 	 &+22 21 03	& 1999 Nov 30, 2000 Nov  8 	& $\alpha$ Aur  	& 6,3     & n,n \\   *
IRAS 04381+2540	 \tablenotemark{a} 	&  04 41 12.48   &  +25 46 37.1 & 2000 Dec 10			& HR 531   	    	& 6       & y   \\   *
IRAS 04489+3042	\tablenotemark{a} 	&  04 52 06.9    &  +30 47 17   & 2000 Feb 21			& HR 1708 	    	& 6       & y   \\   *
GM Aur \tablenotemark{a}          	&04 55 10.2      &+30 21 58     & 2000 Nov  8			& $\alpha$ Tau  	& 3       & n   \\   *
MWC 480          			&04 58 46.27     &+29 50 37.0   & 1999 Nov 30			& $\alpha$ Tau          & 6       & n   \\   *
BN		 			&  05 35 14.17   &  -05 22 23.1 & 2000 Dec 10			& HR 2943           	& 6       & y   \\   *
NGC 2024 IRS2 	 			&  05 41 45.8    &  -01 54 30   & 2000 Dec 10			& HR 2943           	& 6       & y   \\   *
Mon R2 IRS3	 			&  06 07 47.8    &  -06 22 55   & 2000 Dec 10			& HR 2943           	& 6       & y   \\   *
HD 233517        			&08 22 46.71     &+53 04 49.2   & 2000 Nov  8			& $\alpha$ Aur          & 3       & n   \\   *
Hen 3-600A       			&11 10 27.9      &-37 31 52     & 2000 Feb 21			& HR 4786               & 6       & y   \\   *
HD 102647        			&11 49 03.578    &+14 34 19.417 & 2000 Feb 19			& HR 4534               & 6       & y   \\   *
HR 4796A         			&12 36 01.032    &-39 52 10.219 & 2000 Feb 21			& HR 4786               & 6       & y   \\   *
IRAS 14050-4109 \tablenotemark{a} 	&14 05 05.7      &-41 09 40	& 2000 Feb 21			& HR 4786  		& 6       & y   \\   *
HD 163296        			&17 56 21.29     &-21 57 21.9   & 1999 Aug 23, 2000 Jun 20 	& $\beta$ Oph      	& 6,3     & n,y \\   *
HD 179218        			&19 11 11.254    &+15 47 15.630 & 2000 Feb 21			& HR 5908               & 6       & y   \\   *
WW Vul \tablenotemark{a}          	&19 25 58.75     &+21 12 31.3   & 2000 Jun 20			& $\epsilon$ Cyg    	& 3       & y   \\   *
HD 184761        			&19 34 58.97     &+27 13 31.2   & 2000 Nov  8			& $\alpha$ Tau      	& 3       & n   \\   *
HD 216803        			&22 56 24.0529   &-31 33 56.042 & 2000 Dec  9			& HR 1017           	& 6       & y   \\   *
\enddata           
\tablenotetext{a}{Residuals of incomplete subtraction of the 9.7 $\mu$m telluric ozone feature are present in the spectrum.}              
\tablenotetext{b}{Photometric observations only. A 8--13 $\mu$m spectrum was not acquired for this source.}
\end{deluxetable*}

Spectra in the 10 $\mu$m region of a moderate sample of optically visible HAEBE, T Tauri, as well as more heavily embedded YSOs were acquired. The goal was to use the spectra to characterize the composition of the dust, particularly the degree of crystallinity of the silicate grains present. Objects with varying age and mass were chosen in order to put this information in context with the stellar evolutionary process for both low- and high-mass stars. 
The sample includes 7 low-mass class I objects chosen from \citet{kenyon_93} and 3 well known high-mass YSOs: the Becklin Neugebauer object (BN), NGC 2024 IRS2, and Mon R2 IRS3.  These observations probe the composition of largely unprocessed grains and absorption from amorphous silicates is expected. There are ten optically visible class II T Tauri and HAEBE stars in the sample, most of which have very well characterized SEDs.  ISO observations indicate that at least two HAEBE stars in this intermediate-age sample, HD 163296 and HD 179218, possess detectable amounts of crystalline silicates \citep{vandenancker_00, meeus_01}.  
Seven older main sequence stars with debris disks were also included, which possess optically thin IR continua and ages older than 6 Myr. These are the most likely candidates for detection of crystalline silicates, provided warm dust is present.  Our source list and observing dates are provided in Table~\ref{tab:obs}. 

\section{PHOTOMETRIC AND SPECTROSCOPIC OBSERVATIONS}

8--13 $\mu$m spectra/photometry were obtained using the Long Wavelength Spectrometer (LWS) at the W. M. Keck Observatory between 1999 August and 2000 December (see Table~\ref{tab:obs}).  LWS provides diffraction-limited (FWHM = 0.$''$22) imaging (10$''$ field of view, 0.$''$08 pixel$^{-1}$) and long slit (7$''$) spectroscopic (R = 100--1400) capabilities in the 3--25~$\mu$m wavelength range.  Photometry was obtained with the 10.7~$\mu$m filter ($\Delta\lambda$ = 10.0--11.4~$\mu$m) for sources observed during the nights of 2000 Feb 20--21 and December 9, under poor and variable seeing conditions: 0.$''$3--0.$''$6 at 10~$\mu$m. The {\it Nwide} filter and low-resolution grating (LRES) were used to obtain 8--13~$\mu$m spectra, with R = 100, at a dispersion of 0.037~$\mu$m pixel$^{-1}$. Slit widths of 3 and 6 pixels were used, resulting in 0.$''$24 and 0.$''$48 apertures, respectively.  Sources were imaged prior to each spectroscopic observation to ensure optimal placement within the slit. 
For most spectroscopic observations, additional calibration scans were obtained using the Keck routine {\bf LSEC}, in which the data is chopped between an ambient blackbody source (T $\approx$ 274 K) and the sky. 
Darks were taken at the end of each night with exposure times equal to those on source. The raw ``six-dimensional'' LWS data, consisting of images for two chop-nod pairs, were coadded into two-dimensional images, using the IDL routine {\bf LWSCOADD}\footnotemark.  \footnotetext{{\bf LWSCOADD} was written by Gregory D. Wirth and is provided by the observatory at 
http://www2.keck.hawaii.edu/inst/lws/lwscoadd.html.} 

The {\bf IRAF/PHOT} task was used for the photometric data reduction.   
Apertures of radius 0.$''$96 (12 pix) were used for the photometry and the sky background was measured in 2.$''$0--3.$''$2 radius (25--40 pix) annuli. Flat-fielding was found to increase the scatter in the photometry and therefore not performed. 
 A curve-of-growth correction of -0.12 $\pm$ 0.01 mag (based on standard star measurements) to a 1$''$.92 (24 pix diameter) aperture was applied to each star, resulting in magnitudes within $\sim$5$\%$ of the infinite aperture value.  Standard star observations were used to obtain extinction curves for each night. We derived a typical night-to-night variation in the atmospheric zero-points of $\sim$20$\%$ and an average atmospheric extinction correction of -0.32 $\pm$ 0.07 mag airmass$^{-1}$. The root-mean-square scatter in the photometry of the standards is 0.06 mag for 2000 February 20 and 21 and 0.13 mag for 2000 December 9.

The spectroscopic data reduction was performed using the NOAO {\bf IRAF/TWODSPEC and ONEDSPEC} packages.  
After applying pixel masks and flat fields, the images were then divided by those produced from the {\bf LSEC} calibration scan obtained closest in time to the observation, to correct for variations in the spectral response of the grating.  The detector's dark current was found to be negligible over the integration times used and therefore not subtracted from the data. 
Spectral extraction was performed via standard methods using the {\bf IRAF} task $apall$.
For the purposes of wavelength calibration and removal of the telluric ozone absorption feature at $\sim$9.5 $\mu$m, each extracted spectrum was aligned with and divided by a standard star spectrum (see Table~\ref{tab:obs} for details) using the {\bf IRAF} $telluric$ task.  
In some cases (as noted in Table~\ref{tab:obs}), variations in seeing and atmospheric transmission versus wavelength resulted in small differences in the shape of the telluric ozone absorption feature observed in the target and calibrator spectra, and residuals can be seen in the resulting divisions.  After telluric corrections were performed, the resulting spectra were multiplied by a blackbody emission spectrum appropriate to each calibrator to remove any induced slope in the 8--13 $\mu$m region.
Absolute flux calibration was performed using photometric observations at similar wavelengths culled from the literature or obtained in this study (as indicated in Table~\ref{tab:fluxcal}), with scale factors obtained from the spectrum integrated over the photometric bandwidth.
Flux calibrated spectra are shown in Figures~\ref{fig:ysolow}-\ref{fig:debhigh}.  The spectra are overplotted onto the SEDs to indicate the success of this flux calibration method.

\begin{deluxetable*}{lccccc}
\tablewidth{0pt}
\tablecaption{Flux calibration of spectra \label{tab:fluxcal}}
\tablehead{\colhead{} & \colhead{$\lambda$} & \colhead{Flux}  & \colhead{$\Delta{\lambda}$}  & \colhead{} & \colhead{}\\
\colhead{Source} & \colhead{($\mu$m)} & \colhead{(Jy)}  & \colhead{($\mu$m)}  & \colhead{Instrument} & \colhead{Reference}}
\startdata
49 Cet	          			&10.8	& 0.20   $\pm$ 0.04	   & 8.0--13.6	  &KeckII,OSCIR	      & 1 \\ *
HD 17925	  			&10.7	& 0.92   $\pm$ 0.09        &10.0--11.4    &Keck,LWS           & 2 \\ *
HD 22049          			&12	& 9.5    $\pm$ 0.4	   &9.15--14.85	  &IRAS		      & 3 \\ *
IRAS 04016+2610	\tablenotemark{a}  	&10.7	& 2.8    $\pm$ 0.2         &10.0--11.4    &Keck,LWS           & 4 \\ *
		  			&10.3   & 2.5    $\pm$ 0.1	   &9.65--10.95	  &IRTF		      & 5 \\ *
IRAS 04108+2803B \tablenotemark{a} 	&10.7	& 1      $\pm$ 3           &10.0--11.4	  &Keck,LWS           & 4 \\ *
		 			&10.3   & 0.57   $\pm$ 0.09	   &9.65--10.95	  &IRTF		      & 5 \\ *
IRAS 04169+2702				&12.0	& 0.75   $\pm$ 0.05	   &9.15--14.85	  &IRAS 	      & 3 \\ *
IRAS 04181+2654A/B  			&12.0	& 0.36   $\pm$ 0.05	   &9.15--14.85	  &IRAS 	      & 3 \\ *
IRAS 04239+2436	  			&12.0	& 1.71   $\pm$ 0.09        &9.15--14.85	  &IRAS		      & 3 \\ * 
IRAS 04248+2612	  			&10.7	& 0.230  $\pm$ 0.009       &10.0--11.4	  &Keck,LWS           & 4 \\ *
IRAS 04264+2433	  			&10.7	& 0.41   $\pm$ 0.02	   &10.0--11.4	  &Keck,LWS           & 4 \\ *
HARO 6-10A                              &10.0   & 4.3\tablenotemark{b}     & ...    	  &...                & 6 \\ *
HARO 6-10B                              &10.3	& 10	 $\pm$ 2 	   &9.65--10.95	  &IRTF 	      &	5 \\ *
IRAS 04287+1801   			&10.7	& 5.0    $\pm$ 0.4         &10.0--11.4	  &Keck,LWS           & 4 \\ *
IRAS 04295+2251  			&10.7	& 0.69   $\pm$ 0.03	   &10.0--11.4	  &Keck,LWS	      & 4 \\ *
AA Tau	  	  			&10.7	& 0.44   $\pm$ 0.03        &10.0--11.4    &Keck,LWS           & 2 \\ *
IRAS 04325+2402 \tablenotemark{c}   	&10.7 	& 0.060  $\pm$ 0.003       &10.0--11.4    &Keck,LWS           & 4 \\ *
LkCa 15	          			&9.6	& 0.50   $\pm$ 0.15	   & 9.5--9.7	  &ISO,SWS	      & 7 \\ *
IRAS 04381+2540	  			&10.1	& 0.23   $\pm$ 0.05        &9.65--10.95	  &IRTF	              & 5 \\ *
IRAS 04489+3042	  			&10.7	& 0.31   $\pm$ 0.02	   &10.0--11.4	  &Keck,LWS	      & 4 \\ *
GM Aur	          			&10.1	& 0.5    		   &7.55--12.65   &IRTF               & 8 \\ *
MWC 480	          			&9.6	& 8.7    $\pm$ 2.6	   & 9.5--9.7	  &ISO,SWS            & 7 \\ *
BN	          			&10.7	& 88            	   &10.0--11.4	  &Keck,LWS	      & 9 \\ *
NGC 2024 IRS2  	  			&10.5   & 19     $\pm$1            &8.0--13.0     &MANIAC,ESO         & 10  \\ *
Mon R2 IRS3	  			&10.0   & 165    $\pm$5            &9.65--10.95   &IRTF               & 11  \\ *
HD 233517         			&10.1	& 0.439  		   &9.595--10.505 &UKIRT,Berkcam      & 12 \\ *
Hen 3-600A   	  			&10.7	& 0.73   $\pm$ 0.06        &10.0--11.4    &Keck,LWS           & 2 \\ *
HD 102647  	  			&10.7	& 5.51   $\pm$ 0.41        &10.0--11.4    &Keck,LWS           & 2 \\ *
HR 4796A	  			&10.7	& 0.22   $\pm$ 0.02        &10.0--11.4    &Keck,LWS           & 2 \\ *
IRAS 14050-4109   			&10.7   & 0.23   $\pm$ 0.02        &10.0--11.4    &Keck,LWS           & 2 \\ *
HD 163296         			&9.69	& 18.6   $\pm$ 0.5	   &8.84--10.54	  &1m,ESO             &13 \\ *
HD 179218         			&10.7	& 17     $\pm$ 2  	   &10.15--11.25  &O'brien,bolometer  &14 \\ *
WW Vul	          			&9.6	& 2.3    $\pm$ 0.5	   & 9.5--9.7	  &ISO,SWS            & 7 \\ *
HD 184761         			&12	& 2.41   		   &9.15--14.85	  &IRAS		      & 3 \\ *
HD 216803         			&10.7	& 1.53   $\pm$ 0.082       &10.0--11.4    &Keck,LWS           & 2 \\ *
\enddata
\tablenotetext{a}{Photometric errors are larger in the data acquired in this study, therefore 10.3 $\mu$m photometry from (5) were used for flux calibration.}
\tablenotetext{b}{Flux for Haro 6-10B is an estimate by \citet{leinert_89}.}
\tablenotetext{c}{Photometric observations only. A 8--13 $\mu$m spectrum was not acquired for this source.}
\tablerefs{(1) \citealp{jayawardhana_01}; (2) \citealp{metchev_04}; (3) \citealp{IRAS}; (4) this paper; (5) \citealp{myers_87}; (6) extrapolation from \citealp{leinert_89}; (7) \citealp{thi_01};  (8) \citealp{kenyon_93}; (9) Hillenbrand; unpublished, CCD; (10) \citealp{walsh_01}; (11) \citealp{koresko_93}; (12) \citealp{skinner_95}; (13) \citealp{berilli_92};  (14) \citealp{lawrence_90}.}
\end{deluxetable*}

\section{SPECTRAL ENERGY DISTRIBUTIONS (SEDS)}

\begin{figure*}
\includegraphics[width=9.5in]{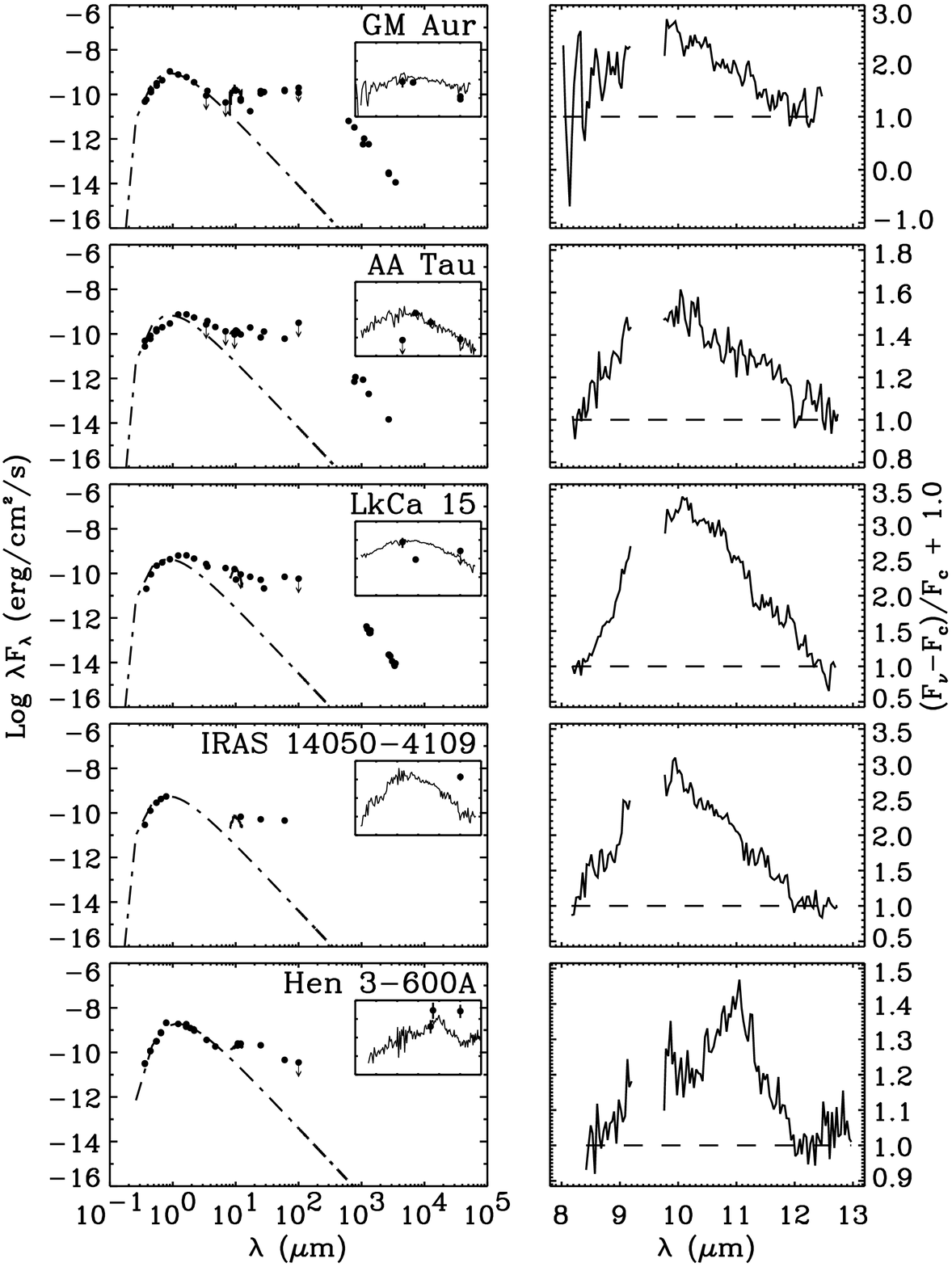}
\vspace{1.0cm}
\figcaption{SEDs and 10 $\mu$m spectra for disks around low-mass T Tauri
stars where the silicate 10 $\mu$m band is in emission.   In the left panel, the SEDs are plotted in units of Log $\lambda$F$_{\lambda}$ overlaid by 
dashed lines representing the stellar photosphere, as fit by the blackbody model described in the text. The inset shows an enlargement of the 8--13 $\mu$m region with linear scaling on both axes. The continuum is obtained by connecting the two endpoints of the spectrum. Continuum subtracted spectra in units of (F$_{\nu}$-F$_{c}$)/F$_{c}$ are shown in the right panel, with the continuum level depicted by dashed lines.
\label{fig:tts}}
\end{figure*}

\begin{figure*}
\includegraphics[width=9.5in]{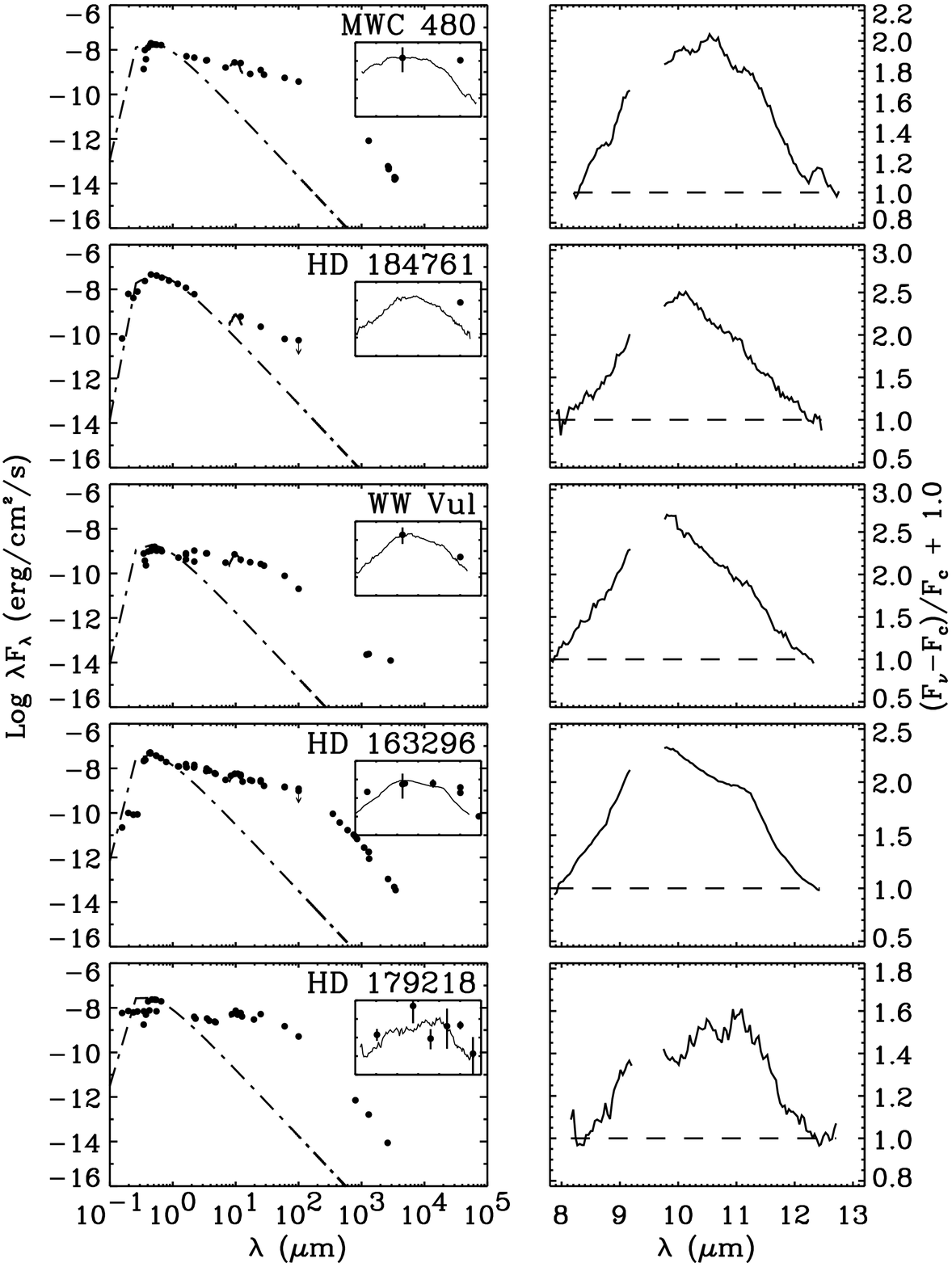}
\vspace{1.0cm}
\figcaption{SEDs and 10 $\mu$m spectra for disks around intermediate-mass 
HAEBE stars where the silicate 10 $\mu$m band is in emission, otherwise the same as Figure~\ref{fig:tts}.
\label{fig:haebe}}
\end{figure*}

\begin{figure}
\includegraphics[width=3in]{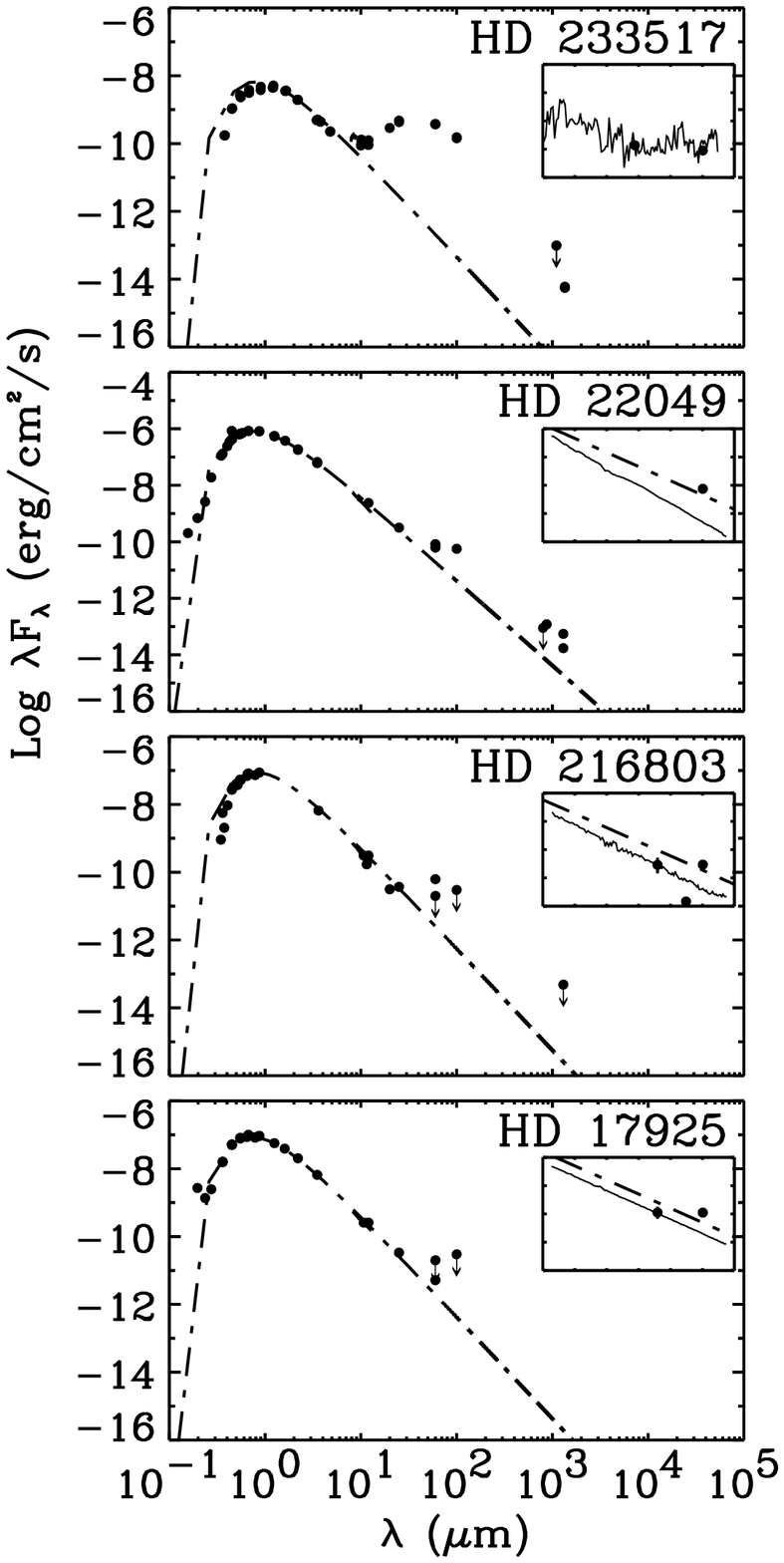}
\figcaption{SEDs including 10 $\mu$m spectra for ``debris'' disks around low-mass stars.  Dashed lines represent the stellar photosphere.  The inset shows an enlargement of the 8--13 $\mu$m region with linear scaling on both axes.
\label{fig:deblow}}
\end{figure}

\begin{figure}
\includegraphics[width=3in]{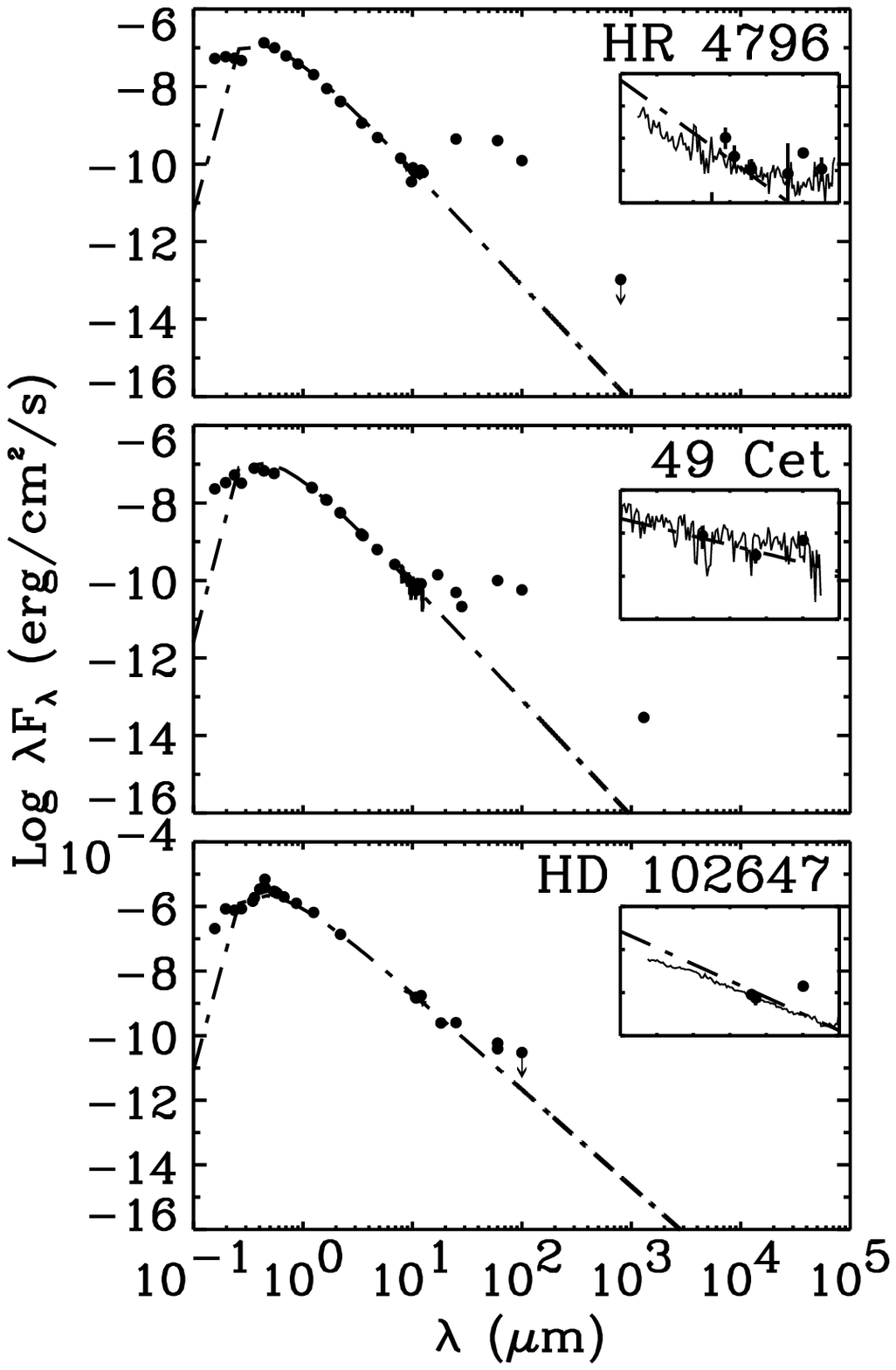}
\figcaption{SEDs including 10 $\mu$m spectra for ``debris'' disks around intermediate-mass stars, otherwise the same as Figure~\ref{fig:deblow}.
\label{fig:debhigh}}
\end{figure}

Continuum fluxes for the sources in our sample were collected from the literature in order to form the spectral energy distributions shown in Figures~\ref{fig:ysolow}-\ref{fig:debhigh}.  The data were not de-reddened, but, as the A$_V$ for all but the class I sources is small ($<3$ mag), this should have a modest effect on fluxes for wavelengths longer than 1.2  $\mu$m (J-band). 
In Figures~\ref{fig:tts}-\ref{fig:debhigh}, the SEDs for stars with known spectral types are overlaid with emission arising from a blackbody at the stellar temperature found in the literature for each source (see Table~\ref{tab:diskparam}) by scaling the peak of the blackbody curve to match the SED at that wavelength. 
More quantitatively, stellar luminosities were found by calculating a color excess and A$_v$ for the given spectral type, applying a bolometric correction to the V or I magnitude (as indicated), and converting to bolometric luminosity using the given distance (see Table~\ref{tab:diskparam} for details).  
For comparison, stellar luminosities from the literature are also presented.  These are similar to the luminosities derived here; a notable exception is HD 179218, for which the V-band magnitudes reported in the literature vary by up to 1.3 magnitudes. For consistency, we will use the luminosities derived here, but note that there may be errors for stars with large variability.

\begin{deluxetable*}{lccccccccccccc}
\tablewidth{0pt}
\tablecaption{Stellar parameters from the literature and calculations from SEDs: revealed disks \label{tab:diskparam}}
\tablehead{\colhead{}  & \colhead{Distance} & \colhead{Spectral} & \colhead{L$_*$} & \colhead{T$_{eff}$} & \colhead{Age} & \colhead{} & \vline & \colhead{} & \colhead{} &\colhead{L$_{*,bol}$\tablenotemark{d}} & \colhead{$\lambda_{onset}$\tablenotemark{e}} & \colhead{}  & \colhead{spectral} \\ 
\colhead{Source} & \colhead{(pc)} & \colhead{Type}  & \colhead{(L$_{\odot}$)} & \colhead{(K)} & \colhead{(Myr)} & \colhead{Ref.\tablenotemark{a}} & \vline & \colhead{Method\tablenotemark{b}} & \colhead{A$_v$\tablenotemark{c}} & \colhead{(L$_{\odot}$)}    & \colhead{($\mu$m)}  & \colhead{L$_{IR}$/L$_*$\tablenotemark{f}} & \colhead{index\tablenotemark{g}}} 
\startdata
49 Cet          &61                &A1V     &23.44  &9500   &7.8   &1,2  &\vline &V/B-V &0.21 & 22  &25   &  1.1(-3)  & 1.94 \\ 
HD 17925        &10.4              &K1V     &...    &5000   &80    &3    &\vline &V/B-V &0    &0.43 &60   &$<$6.0(-5)    & 2.64 \\ 
HD 22049        &3.2               &K2V     &...    &5000   &330   &3    &\vline &V/B-V &0    &0.33 &60   &9.7(-5)    & 2.61 \\ 
AA Tau          &140               &K7      &0.71   &4060   &2.4   &2,4  &\vline &I/V-I &1.0  &0.60 &2.2  &0.48       & 0.84 \\ 
LkCa 15         &140               &K5:V    &0.74   &4350   &2.0   &2,5  &\vline &I/V-I &1.1  &0.86 &1.2  &0.49       & 0.90 \\ 
GM Aur          &140               &K7      &0.724  &4060   &1.8   &2,4  &\vline &J/J-H\tablenotemark{h} &0.06  &0.73  &6.9  &0.43    & 0.48 \\ 
MWC 480\tablenotemark{i}         &131--140 &A3e     &11--32   &8460--8890   &4.6   &6,7    &\vline &V/B-V &0.25 & 14  &1.6  &0.29       & 0.53 \\ 
HD 233517      &23                &K2      &0.20   &4500   &...   &8    &\vline &I/V-I &0.30 &0.12 &10   &0.12       &0.62 \\ 
Hen 3-600A      &50.0              &M4      &...    &3290   &10.0  &9   &\vline &I/V-I &0.10 &0.43 &11   &9.1(-3)    &0.62 \\ 
HD 102647       &11.1              &A3V     &...    &8590   &240   &3    &\vline &V/B-V &0    & 14  &60   &$<$3.9(-6)    & 2.59 \\ 
HR 4796A        &76                &A0V     &35     &10000  &8.0   &10,11 &\vline &I/V-I &0    & 26  &25   &4.6(-3)    &0.92 \\ 
IRAS 14050-4109	&140               &K5      &...    &4405   &...   &12   &\vline &I/V-I &0.81 &0.78 &12   &8.6(-2)~\tablenotemark{j}  &0.68    \\
HD 163296       &122$^{+17}_{-13}$ &A1Ve    &30.2   &9332   &4.0   &6    &\vline &V/B-V &0.34 & 32  &1.2  &0.28       & 0.58 \\ 
HD 179218       &240$^{+70}_{-40}$ &B9e     &316    &10471  &0.1   &6    &\vline &V/B-V &0.59 & 110 &2.2  &0.32       & 0.13 \\ 
WW Vul          &550               &A3e/B9V &43     &8600   &...   &14   &\vline &V/B-V &0.95 & 30  &1.6  &0.50       & 0.57 \\ 
HD 184761       &65                &A8V     &...    &7500   &...   &15   &\vline &V/B-V &0.09 & 7.4 &2.2  &0.10       &  1.38 \\ 
HD 216803       &7.70              &K5Ve    &...    &4555   &200  &16,17 &\vline &V/B-V &0.06 &0.20 &60   &$<$7.0(-4)    &  2.67 \\ 
\enddata  
\tablecomments{Parameters on the left are from the literature as noted; those on the right are calculated here.}
\tablenotetext{a}{Where two references are listed, the second is for the age.}
\tablenotetext{b}{Method is in the form of A/B-C, where B-C is the color used to calculate the stellar A$_v$ and A is the band to which the bolometric correction was applied.}
\tablenotetext{c}{A$_v$ were calculated from the spectral type given and color excess E(B-C), as defined in the Method column.}
\tablenotetext{d}{Stellar bolometric luminosities were calculated by applying a bolometric correction to the magnitude of band A (defined in the Method column) and converting to luminosity using the given distance.}
\tablenotetext{e}{$\lambda_{onset}$ is the wavelength at which the stellar blackbody and a polynomial fit to the SED diverge.}
\tablenotetext{f}{ L$_{IR}$ is calculated by integrating the SED between $\lambda_{onset}$ and 1 cm  and subtracting the integral of the stellar blackbody over the same region. L$_{*}= $L$_{*,bol}$.}
\tablenotetext{g}{The spectral indices are calculated following the method of \citet{kenyon_h95}, $spectral\;index = - \frac{Log (\lambda_b F_{\lambda_{b}}) - Log (\lambda_a F_{\lambda_{a}})}{Log (\lambda_b) - Log (\lambda_a)}$. $\lambda_a\approx2.2$ $\mu$m for most sources, excepting IRAS 14050-4109 (0.8 $\mu$m) and HD 216803 (3.6 $\mu$m); $\lambda_b=25$ $\mu$m for all sources.}
\tablenotetext{h}{J- and H-band data was used to calculate the luminosity for GM Aur, as the I-band magnitude seemed anomalously high.} 
\tablenotetext{i}{d $=131$ pc and T$_{eff}=8710$ were used for the luminosity calculations for MWC 480.}
\tablenotetext{j}{The SED for IRAS 14050-4109 could not be extrapolated beyond the last observed point (100$\mu$m) and therefore this luminosity ratio is a lower limit.}
\tablerefs{(1) \citealp{coulson_w98}; (2) \citealp{thi_01}; (3) \citealp{habing_d01};  (4) \citealp{hartmann_c98}; (5) \citealp{strom_s89}; (6) \citealp{vandenancker_d98}; (7) \citealp{simon_d01}; (8) \citealp{skinner_95}; (9) \citealp{torres_03}; (10) \citealp{jura_z93}; (11) \citealp{stauffer_95}; (12) \citealp{gregorio-hetem_02}; (13) \citealp{mannings_s00}; (14) \citealp{natta_p01}; (15) \citealp{miroshnichenko_m99}; (16) \citealp{barradoynavascues_97}, (17) \citealp{santos_04}.} 
\end{deluxetable*}

\begin{deluxetable}{lcccc}
\tablewidth{0pt}
\tablecaption{Calculations from SEDs: embedded objects \label{tab:irasparam}}
\tablehead{\colhead{}  & \colhead{} & \colhead{Distance} &\colhead{} & \colhead{Spectral} \\
\colhead{Source}  & \colhead{L$_{tot}$} & \colhead{(pc)} &\colhead{Reference} & \colhead{Index\tablenotemark{a}}} 
\startdata	                                                
IRAS 04016+2610   	&  3.7 &140 &1     & -1.03            \\
IRAS 04108+2803B  	& 0.75 &140 &1     & -1.09            \\
IRAS 04169+2702		& 1.20 &140 &3     & -1.29            \\
IRAS 04181+2654A   	& 0.34 &140 &1,2   & -0.31            \\
IRAS 04181+2654B   	& 0.34 &140 &1,2   & -0.44            \\
IRAS 04239+2436   	& 1.39 &140 &1     & -1.10            \\
IRAS 04248+2612   	& 0.10 &140 &1     & -0.49            \\
IRAS 04264+2433   	& 0.24 &140 &1     & -1.00            \\
Haro 6-10A	        & 6.5  &140 &3     & -0.71            \\
Haro 6-10B	        & 6.5  &140 &3     & -1.56            \\
IRAS 04287+1801   	& 0.29 &140 &1     & -1.78            \\
IRAS 04295+2251  	& 0.41 &140 &1,2   & -0.23            \\
IRAS 04381+2540   	& 0.70 &140 &2     & -1.31            \\
IRAS 04489+3042   	& 0.31 &140 &1     & -0.31            \\
BN 	  	  	&  980 &450 &4     & -1.25            \\
NGC 2024 IRS2       	& 5800 &415 &5     & -1.74            \\
Mon R2 IRS3   	  	&30000 &830 &6     & -2.77            \\
\enddata    	
\tablecomments{The luminosities for several close sources are those of the combined sources due to the large IRAS beam.}
\tablenotetext{a}{The spectral indices are calculated as in Table~\ref{tab:diskparam}. $\lambda_a=2.2$ $\mu$m for all sources; $\lambda_b=25$ $\mu$m for all low-mass stars and  20, 25, and 39 $\mu$m for BN, NGC 2024 IRS2, and Mon R2 IRS3, respectively.}
\tablerefs{(1) \citealp{hartmann_02}; (2) \citealp{IRAS}; (3) \citealp{myers_87}.; (4) \citealp{gezari_98}; (5) \citealp{haisch_l01}; (6) \citealp{preibisch_02}.}
\end{deluxetable}  

The IR luminosity and the ratio L$_{IR}$/L$_{*}$ (indicative of dust geometry for optically thick disks and of midplane optical depth for optically thin disks) were estimated from the SEDs collected for our sample in the following manner. First, the wavelength ($\lambda_{onset}$) at which the photometric flux is greater than twice that of the stellar blackbody was evaluated\footnotemark. \footnotetext{$\lambda_{onset}$ is calculated to give the reader an impression of where the SED begins to deviate from photospheric and was chosen to be the point at which the photometric flux is greater than {\it twice} that of the stellar blackbody as this minimizes the errors due to variability in the photometric measurements.} The infrared luminosity of the disk L$_{IR}$ was then calculated by integrating the SED between $\lambda_{onset}$ and 1 cm  and subtracting the integral of the stellar blackbody over the same region.  The resulting IR-to-stellar luminosity ratios (L$_{IR}$/L$_{*}$), as well as the wavelengths of the onset of excess radiation ($\lambda_{onset}$), are presented in Table~\ref{tab:diskparam}. As the spectral types and A$_V$ for the embedded YSOs are uncertain, only total luminosities were calculated; resulting values of L$_{tot}$ are presented in Table~\ref{tab:irasparam}.

The SEDs were used to establish the continuum around the 8--13~$\mu$m silicate emission feature, which is extremely important for analysis of the feature shape and the silicate composition. To establish the continuum for HAEBE stars, \citet{vanboekel_03} fit each SED with a blackbody for the star and several blackbodies of different temperatures for the dust emission and interpolated between the two.  \citet{natta_00} fit a powerlaw to the edges of the silicate emission feature to remove the continuum in a sample of T Tauri stars. Upon examination of the overlays of the spectra on the SEDs in Figure~\ref{fig:tts}-\ref{fig:haebe}, we find that for our sample the continuum is adequately represented by  connecting the two endpoints of the spectrum (averaging over 0.20 $\mu$m near 8.3 and 12.2 $\mu$m). 

\begin{figure*}
\includegraphics[width=9.5in]{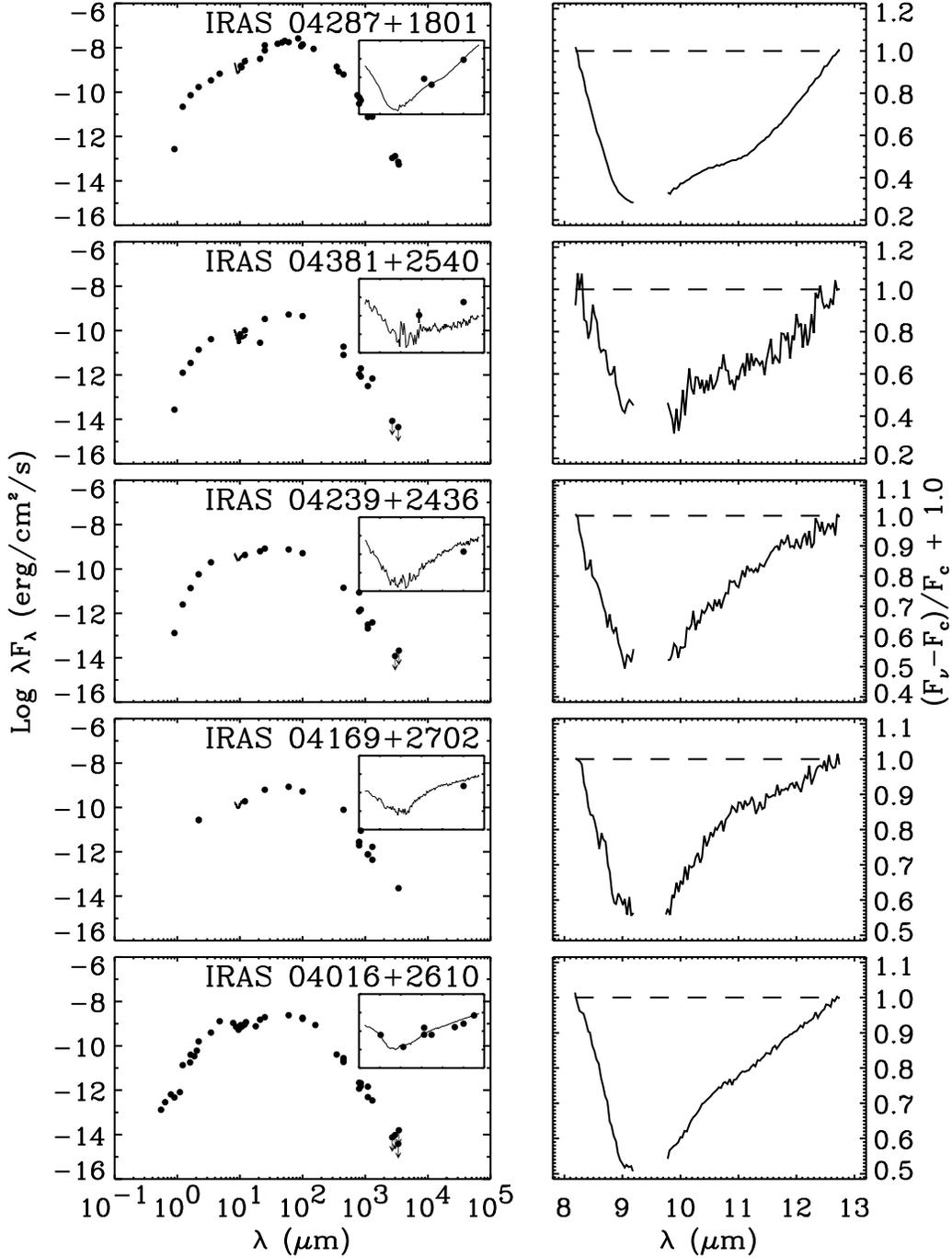}
\vspace{1.0cm}
\figcaption{SEDs and 10 $\mu$m spectra for low-mass stars where silicates are in absorption.
The left panel displays the SEDs for each source (see appendix A for references) in units of Log $\lambda$F$_{\lambda}$, with the 10 $\mu$m spectra included for reference. The insets show enlargements of the 8--13 $\mu$m regions with linear scaling on both axes.  Open symbols are data for which a single source within a binary or multiple can not be distinguished.
The continuum is in most (but not all) cases adequately represented by  connecting the two endpoints of the spectrum and normalized spectra are shown in the right panel in units of (F$_{\nu}$-F$_{c}$)/F$_{c}$ (see text for details). The dashed line depicts the continuum level.
\label{fig:ysolow}}
\end{figure*}

\begin{figure*}
\addtocounter{figure}{-1}
\includegraphics[width=9.5in]{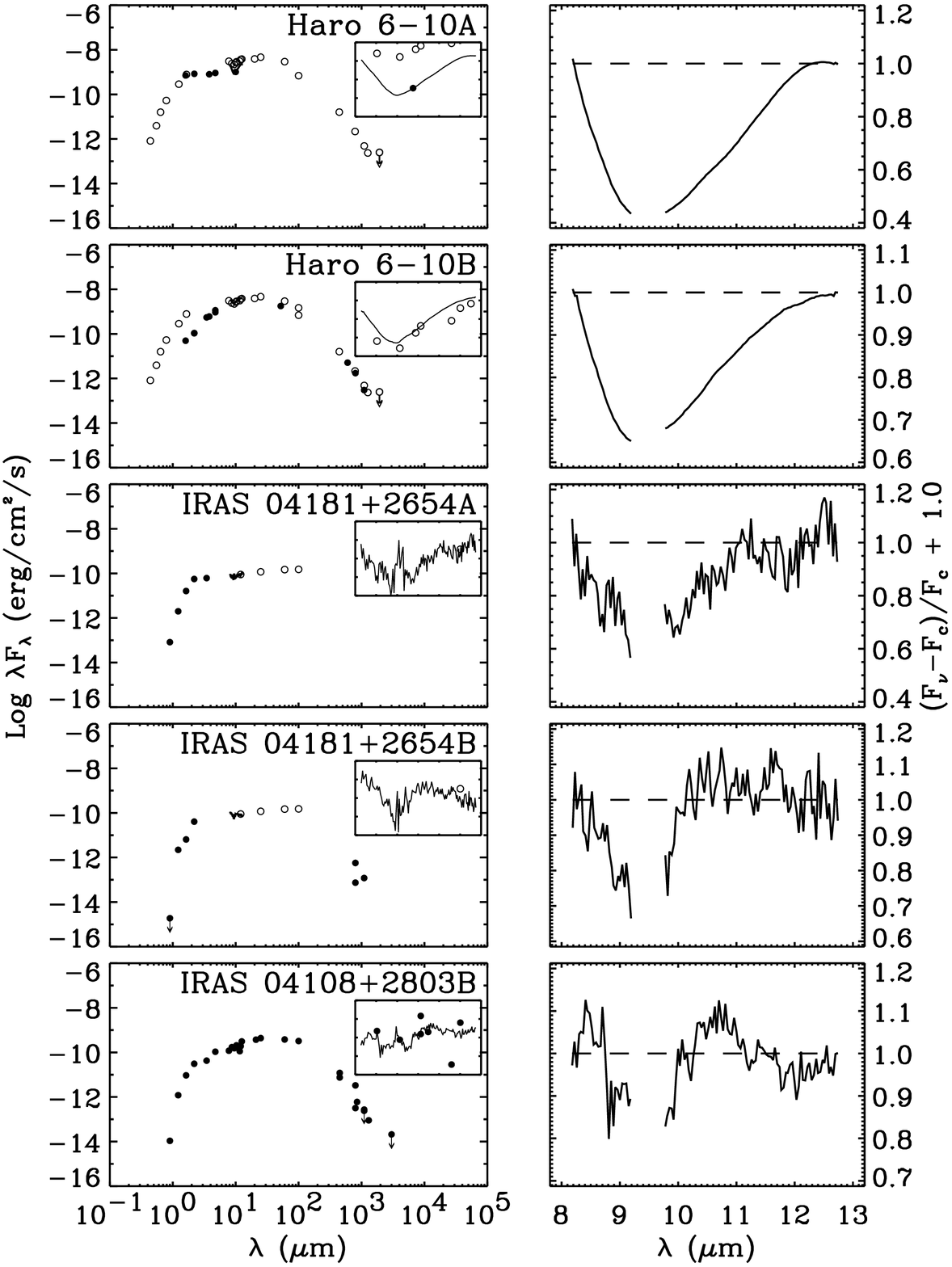}
\vspace{1.0cm}
\figcaption[ ]{-continued}
\end{figure*}

\begin{figure*}
\addtocounter{figure}{-1}
\includegraphics[width=9.5in]{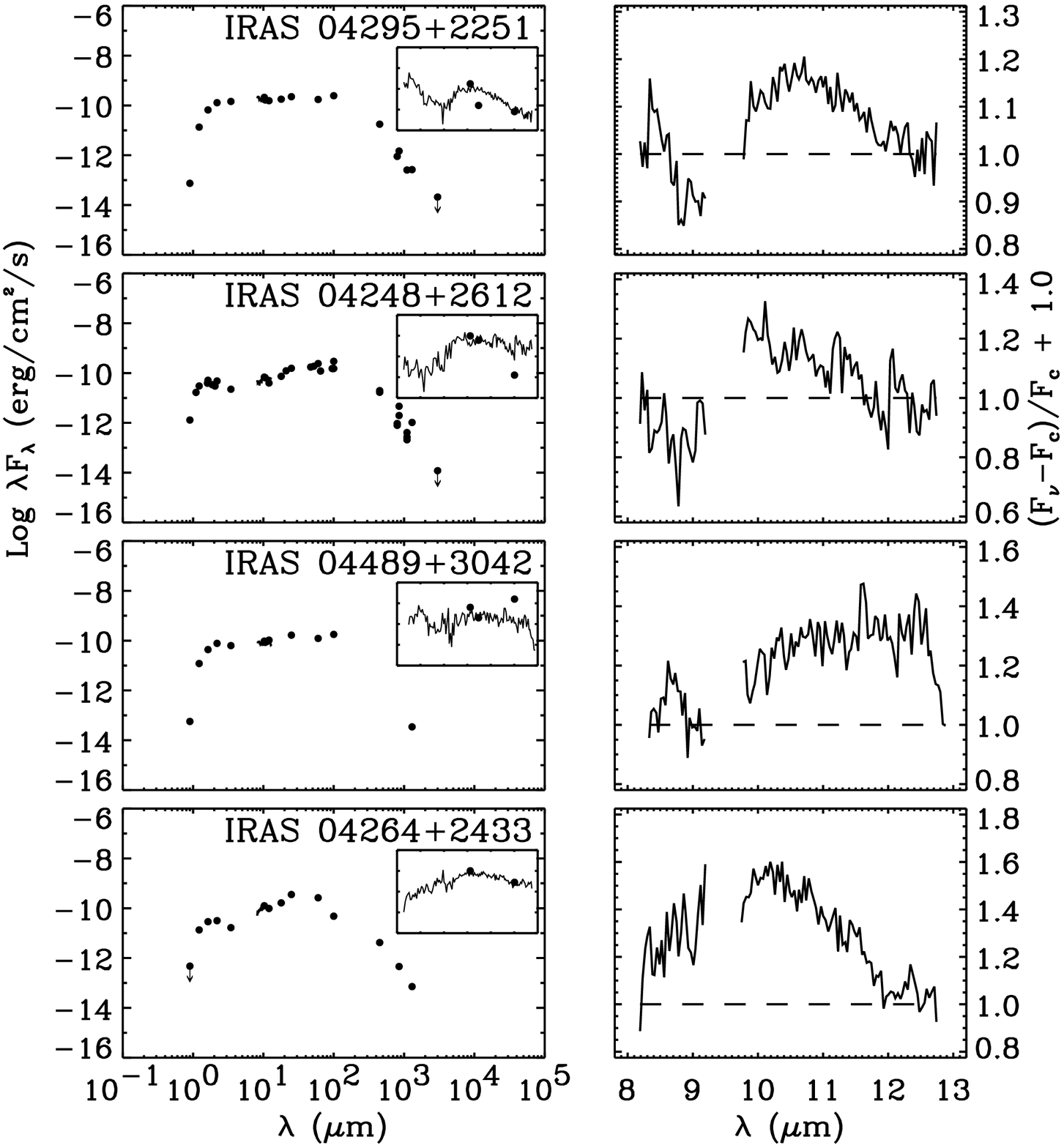}
\figcaption[ ]{-continued}
\end{figure*}

\begin{figure*}
\includegraphics[width=9.5in]{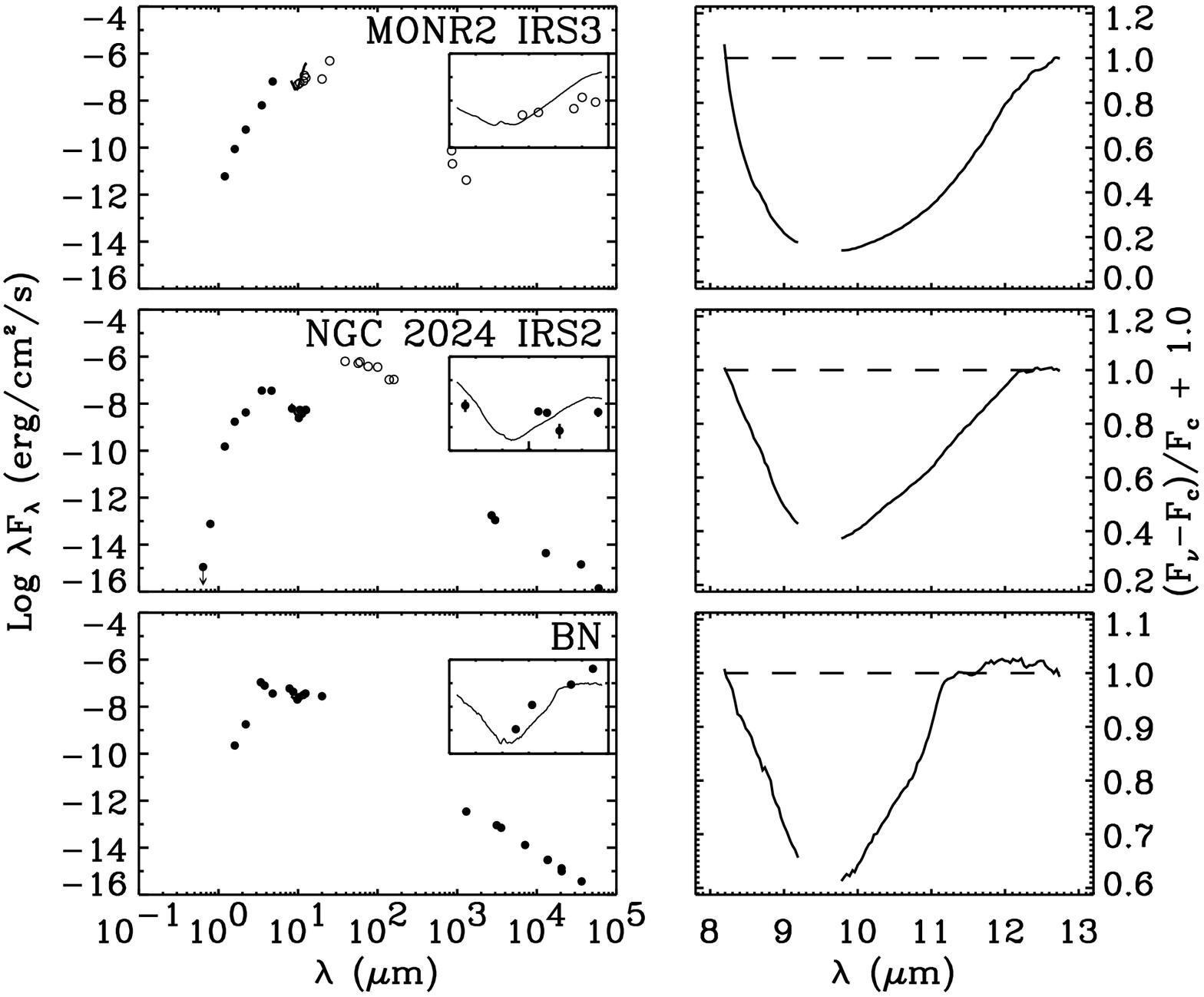}
\figcaption{SEDs and 10 $\mu$m spectra 
for high mass stars where silicates are in absorption, otherwise the same as Figure~\ref{fig:ysolow}.
\label{fig:ysohigh}}
\end{figure*}

In general, we would expect the shape of the 8--13 $\mu$m feature to be related to the SED and to the relative disk versus stellar radiation.  Specifically, if changes in the silicate feature are due to processing of the dust, through growth and/or crystallization, then the shape of the silicate emission feature should be related to the evolutionary stage of the source.  Because the spectral index is an indicator of the evolutionary class, we expect that it may be related to the shape of the 8--13 $\mu$m spectra.  In the following sections we explore these trends, first through qualitative discussion and then through quantitative characterization of the spectral strength and shape.

\section{DISCUSSION OF THE 10 $\mu$m SPECTRA}

The flux calibrated 0.1-10$^5$ $\mu$m SEDs and 8--13 $\mu$m spectra are shown in Figures~\ref{fig:ysolow}-\ref{fig:debhigh}, with the continuum subtracted silicate features shown in the right column in Figures~\ref{fig:ysolow}--\ref{fig:haebe}. The sources are placed into categories based on the shape of the SEDs and 8--13 $\mu$m spectra. In general, the debris disks are found to have little or no evidence of silicate emission, the T Tauri and HAEBE sources show strong to moderate silicate emission and the embedded sources show silicate absorption, or a combination of absorption and emission.  
Possible evolutionary trends are shown in Figures~\ref{fig:evoltts} and \ref{fig:evolhae} for the low-mass and intermediate-mass stars, respectively.  The trend shown in Figure~\ref{fig:evolhae} is similar to that suggested for HAEBE stars by \citet{meeus_99}.  The evolution of low-mass star spectra and SEDs does not appear to be significantly different from intermediate-mass stars, with embedded sources showing amorphous olivine in absorption, followed by stages of dust processing (including grain growth and/or silicate crystallization) resulting in variations in the shape of the emission feature, and finally photospheric spectra for debris disks.  One possible difference between the low- and intermediate-mass stars in our sample is that for the intermediate-mass stars, crystalline silicates have been found in sources which possess large near- and far-IR excesses, but the one low-mass star toward which crystalline silicates have been found, Hen 3-600A \citep[see also ][]{honda_03,uchida_04}, has an SED with no near-IR excess and diminishing far-IR flux.  There have been far fewer detections of crystalline silicates toward low-mass, with respect to high-mass, stars reported in the literature as well.  It is not at all clear, however, that the surveys to date are drawn from representative samples.

\subsection{Silicate Absorption Spectra}

The silicate feature is clearly seen in absorption in 9 low-mass embedded
sources and in all 3 high mass protostars in our sample 
(see Figures~\ref{fig:ysolow} and \ref{fig:ysohigh}).
The spectra of IRAS 04287+1801 and IRAS 04381+2540 (and possibly Mon R2 IRS3) 
possess a shallow absorption feature near 11.2 $\mu$m in addition to the 
9.7 $\mu$m absorption seen toward the other embedded sources.
The additional absorption feature has rarely been seen \citep[e.g.,][]{boogert_p04}, 
 and its origin may be related to grain mantle ice constituents and not silicates 
(Boogert, private communication).
In 5 other low-mass embedded objects the silicate feature appears with 
absorption superposed on emission between 
9 and 12 $\mu$m.  In the most extreme case, IRAS 04264+2433, 
the spectrum shows almost pure silicate emission.  
These spectra may be explained by a source geometry in which 
cold dust located along the line of sight is absorbing against silicate emission from 
warmer dust located near the protostar.  Radiative transfer modeling is required to explore this
scenario and constrain whether the cold dust is in fact associated with these sources or associated with  remnant unbound material in the dense core.  

The absorption spectra 
exhibit a variety of morphologies and have been ordered in Figure~\ref{fig:ysolow} roughly according the strength of the absorption feature.
This may correspond to an evolutionary sequence from strong
 ISM-like absorption through superposed absorption+emission and will be discussed more quantitatively in $\S$6.1.  

\subsection{Silicate Emission Spectra}

The silicate feature is seen in emission toward 5 T Tauri stars and 5 HAEBE stars in our sample (see Figures~\ref{fig:tts} and \ref{fig:haebe}).  8 out of 10 of these stars possess significant IR luminosities (L$_{IR}$/L$_*$ $>$ 0.1), with the exception of IRAS 14050-4109 and Hen 3-600A (L$_{IR}$/L$_*$ $\sim$ 10$^{-2}$, similar to debris disks).  The shape of the emission feature varies from smooth single peaks near 9.7 $\mu$m, consistent with amorphous olivine, to more complex spectra with additional peaks near 10.2 and 11.3 $\mu$m, indicative of enstatite and forsterite components.  This suggests that silicate processing is occurring within disks around optically revealed T Tauri and HAEBE stars.

8--13 $\mu$m spectra similar to that of amorphous olivine were observed toward 4/5 T Tauri stars (GM Aur, AA Tau, IRAS 14050-4109  and LkCa 15).  The emission features are wider in comparison to absorption features in the ISM.  This could either be due to the addition of $\sim$10.2--11.3 $\mu$m emission from crystalline silicates or broadening of the amorphous emission feature due to increased grain sizes. The absence of a strong crystalline component in the emission spectra indicates that the wider feature is more likely related to grain growth (discussed below) than to crystallization. 
Crystalline silicates appear to be absent even in the two sources with 
evidence of disk clearing, 
IRAS 14050-4109 (with L$_{IR}$/L$_*$ $\sim$ 10$^{-2}$) and GM Aur \citep[possessing a gap at small radii,][]{rice_w03}.

Possible evidence for crystallization appears in the spectra observed toward 4/5 HAEBE stars (MWC 480, HD 163296, HD 184761, and WW Vul).   HD 163296, HD 184761, and WW Vul each possess an emission feature characteristic of amorphous olivine as well as a small peak at 11.3 $\mu$m, consistent with crystalline forsterite \citep[though this peak is much weaker than that seen for $\beta$ Pictoris or HD 100546,][]{knacke_f93, meeus_01}. MWC 480 possesses a peak near 10.5 $\mu$m, which is consistent with crystalline enstatite.  
The three disks with 11.2 $\mu$m peaks are young intermediate-mass stars 
($\sim$2 M$_{\odot}$) with SEDs depicting substantial IR emission from 
warm dust, indicating that the component producing the feature must be present 
at a young evolutionary stage.
Analysis of the crystalline silicate content in these disks, however, may be
complicated by the 11.2 $\mu$m PAH emission band.  
Clear identification of the 11.3 $\mu$m feature as arising from crystalline 
silicates can only be obtained if there is evidence of their longer wavelength features, or if the presence of PAHs can be ruled out.
HD 163296 and WW Vul have been observed with ISO and no evidence of PAH emission in the 6--12 $\mu$m region was detected \citep{meeus_01,acke_v04}; longer wavelength crystalline silicate features were detected toward HD 163296.  In the same study, PAH emission at 6.2 $\mu$m was detected toward MWC 480, but similar features at 7.6 and 8.5 were not.  No previous observations of  HD 184761 are available.

Finally, one low-mass T Tauri star (Hen 3-600A) and one HAEBE star (HD 179218), 10 Myr and 6 Myr of age, respectively, each show a strong peak at 11.3 $\mu$m that dominates the spectrum.  The shape of the emission feature is consistent with that of crystalline forsterite (including the correlating small peak at 10.2 $\mu$m).  These spectra are similar to that of the first widely recognized crystalline silicate source, HD 100546 \citep{bouwman_03}.  In contrast to the detection of crystalline silicates toward several HAEBE stars, to date crystalline silicates have been detected toward only a small number of $\sim$solar-mass stars (TW Hya, \citealp{sitko_00}; VW Cha and Glass I, \citealp{meeus_03}).
Our sample seems consistent with this trend, with 1/5 low-mass star disks and 4/5 HAEBE star disks showing signs of crystalline silicates.  Our sample is small, however, and may not be representative.  Observations of a larger sample of low-mass stars are needed to quantify and understand the differences in the composition of dust around low- versus intermediate-mass stars.

\subsection{Featureless Spectra}

The spectra and SEDs for debris disks around low- and intermediate-mass stars are shown in Figures~\ref{fig:deblow} and \ref{fig:debhigh}, respectively.  
No silicate emission was seen toward these disks --- their spectra are predominantly photospheric, with a few exhibiting dust continua.  
The lack of silicate emission may indicate either a paucity of small grains ($a<10~\mu$m), or a dearth of optically thin material inside of $\sim$10 AU (T~$\geq$~200 K). The absence of a substantial continuum at 10 $\mu$m is consistent with the SEDs for HD 102647, HD 17925, HD 22049 and  HD 216803, which show very low IR and millimeter excesses (L$_{IR}/$L$_{*}<10^{-3}$) with $\lambda_{onset}\ge 60$ $\mu$m. This suggests that the debris disks  have lost most of their (warm, small) circumstellar silicate grains.  
49 Ceti, HR 4796A, and HD 233517\footnotemark, however, still possess substantial far-IR and millimeter excesses. \footnotetext{There is some evidence that HD 233517 may be a Li-rich giant and not a young star with a debris disk. See \citet{balachandran_00} for details.}
This substantial amount of cold dust combined with the absence of silicate emission suggest the possible existence of large gaps in these disks at radii close to the star, 
similar to those modeled in the debris disks around $\epsilon$ Eri, Vega, and HR 4796A \citep{ozernoy_00, wyatt_99, wyatt_03}.  

\subsection{Evolution of Silicate Spectra}

\begin{figure*}
\hspace{-3.3cm}
\includegraphics[width=9.6in]{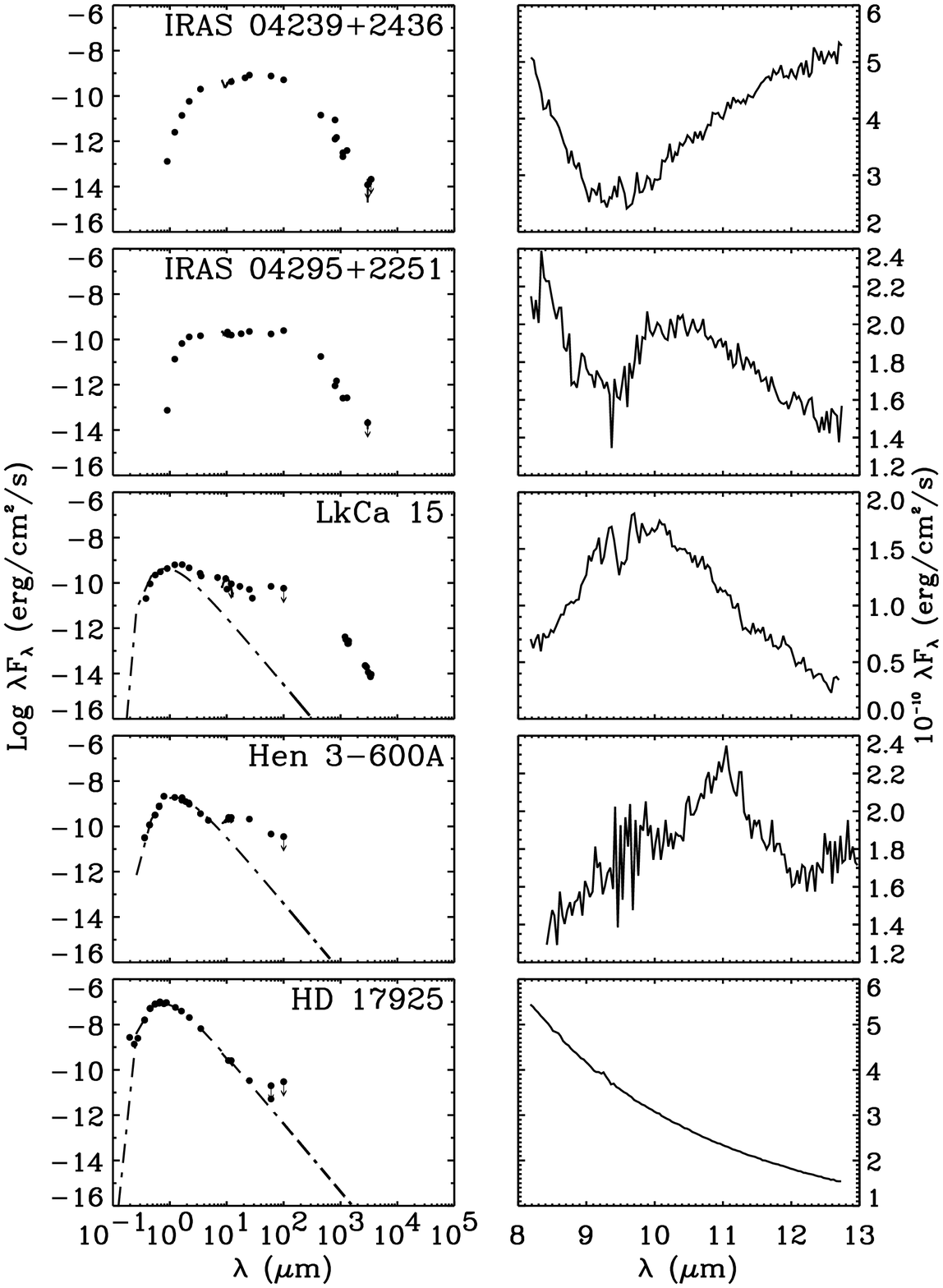}
\vspace{1.0cm}
\figcaption{Evolution of the silicate spectrum for low-mass stars.  The left panel displays the SEDs for selected sources in units of Log $\lambda$F$_{\lambda}$, with the 10 $\mu$m spectra included for reference. The right panel shows an enlargement of the 8--13 $\mu$m region with linear scaling on both axes.  Symbols and lines are as described in Figures~\ref{fig:ysolow}-\ref{fig:debhigh}. The evolutionary sequence for low-mass stars is quite similar to that for intermediate-mass stars \citep{meeus_99}.
\label{fig:evoltts}}
\end{figure*}

\begin{figure*}
\hspace{-3.3cm}
\includegraphics[width=9.6in]{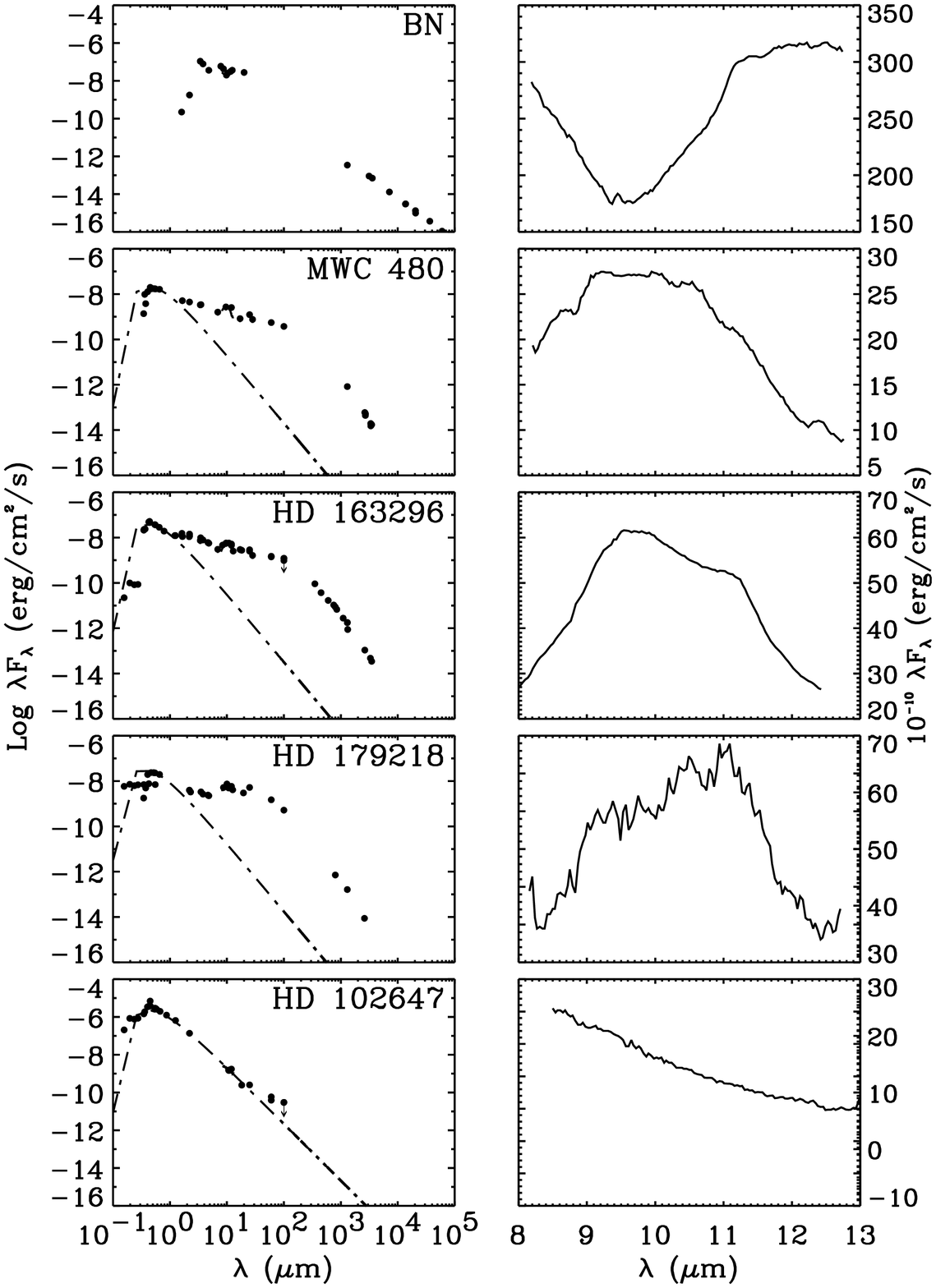}
\vspace{1.0cm}
\figcaption{Evolution of the silicate spectrum for intermediate-mass stars. Otherwise the same as Figure~\ref{fig:evoltts}. The evolutionary sequence is similar to that observed for intermediate-mass stars in ISO studies \citep{meeus_99}.
\label{fig:evolhae}}
\end{figure*}

The SEDs and silicate spectra  of the sample indicate a possible evolutionary trend; silicate  absorption is observed toward embedded protostars, strong silicate emission arises from objects with optically thick protoplanetary disks and nearly photospheric spectra are observed toward optically thin debris disks.  The details of this trend are discussed below for both low- and intermediate-mass stars using a representative subset of our sample as shown in Figures~\ref{fig:evoltts} and \ref{fig:evolhae}.

Figure~\ref{fig:evoltts} shows the evolutionary sequence for material around low-mass stars.  The SED for IRAS 04239+2438 indicates an embedded protostar, and the 8--13 $\mu$m spectrum depicts deep silicate absorption (similar to 7 other embedded sources in our sample).  The SED of LkCa 15 is fit by an optically revealed star and near- to far-IR excess from an optically thick protoplanetary disk \citep[c.f.,][]{chiang_j01}, and strong silicate emission is observed toward this source. Similar SEDs and spectra are seen for 3 other sources.  The mid-IR spectrum of IRAS 04295+2251 shows contributions from both silicate emission, peaking near 10.5 $\mu$m, and silicate absorption near 9.5 $\mu$m and the SED suggests that this source (and 5 others) may be intermediary between IRAS 04239+2438 and LkCa 15, although the exact geometry is not yet well understood.
As discussed above, the SED of Hen 3-600A shows no near-IR excess and very little far-IR flux and silicate emission peaking characteristic of forsterite (peaking near 11.3 $\mu$m), indicating substantial dust processing in this disk.  Finally, HD 17925 is characteristic of 4 sources in our sample with SEDs consistent with a stellar photospheres (out to 60 $\mu$m) toward which no silicate feature is observed, indicating an absence of warm, small circumstellar grains, and longer wavelength excess.

A similar evolutionary sequence is shown for the intermediate-mass stars (Figure~\ref{fig:evolhae}), from silicate absorption toward the fully embedded Becklin Neugebauer object (BN) to silicate emission in the three optically thick disks and a nearly photospheric spectrum toward the optically thin disk encircling HD 102647.  No sources with absorption+emission spectra were seen for the intermediate-mass stars. The sample, however, consisted of only three embedded protostars and there are large differences in morphology for these absorption spectra.  Between the spectra of MWC 480, HD 163296 and HD 179218, the silicate emission feature seems to evolve from amorphous olivine peaking near 9.5 $\mu$m to crystalline forsterite, peaking at 11.3 $\mu$m, suggesting dust processing, but there appears to be no corresponding evolution of the SEDs.  Furthermore, MWC 480 shows silicate emission peaking near 10.5 $\mu$m, indicative of crystalline enstatite, not forsterite.

\section{QUANTIFYING SILICATE SPECTRA}

Having presented our spectra and discussed their overall properties qualitatively, we now embark on a quantitative analysis of the spectra.  
Detailed interpretation of silicate emission/absorption features necessitates an understanding of the optical depth of the circumstellar envelope or the vertical structure of the circumstellar disk.  
Because our sample encompasses a variety of complex circumstellar environments ranging from deeply self-embedded young stars to optically thin debris disks, such detailed modeling will not be presented here.
Instead, we investigate empirical correlations between several
properties of the silicate features and attempt to relate the observed trends to circumstellar properties and dust size and composition.
In $\S6.1$, we begin by evaluating the impact of the source morphology on the structure of the observed silicate feature by investigating correlations between the spectral shape and SED properties.  
Next, we explore the relationship between the FWHM of the 8-13 $\mu$m emission/absorption feature and the central wavelength of the peak/dip ($\S$6.2), which are connected to optical depth and dust composition.
The relationship between the 9.8 and 11.3 $\mu$m fluxes and the peak-to-continuum ratio is indicative of grain size as well as crystallization fraction, as demonstrated by previous studies of HAEBE and T Tauri stars  \citep{przygodda_03, vanboekel_03}.  Thus, relationships between spectral shape and dust composition and grain size will be explored in $\S$6.3--6.4.
By applying these methods of analysis to our spectra, we hope to reveal more subtle evolutionary trends than are apparent from simple visual inspection of the data.

\subsection{Silicate feature strength and SED Properties}

Based on Figures~\ref{fig:evoltts} and \ref{fig:evolhae} we have argued for evidence of an evolutionary sequence in the 10 $\mu$m feature from absorption to mixed absorption and emission to pure emission, which may correspond to an evolutionary sequence in SED morphology.   To explore this relationship more quantitatively we investigate possible correlations between 10 $\mu$m spectral properties and SED slope or IR excess luminosity.

\begin{figure}[ht]
\includegraphics[angle=90,width=9.5cm]{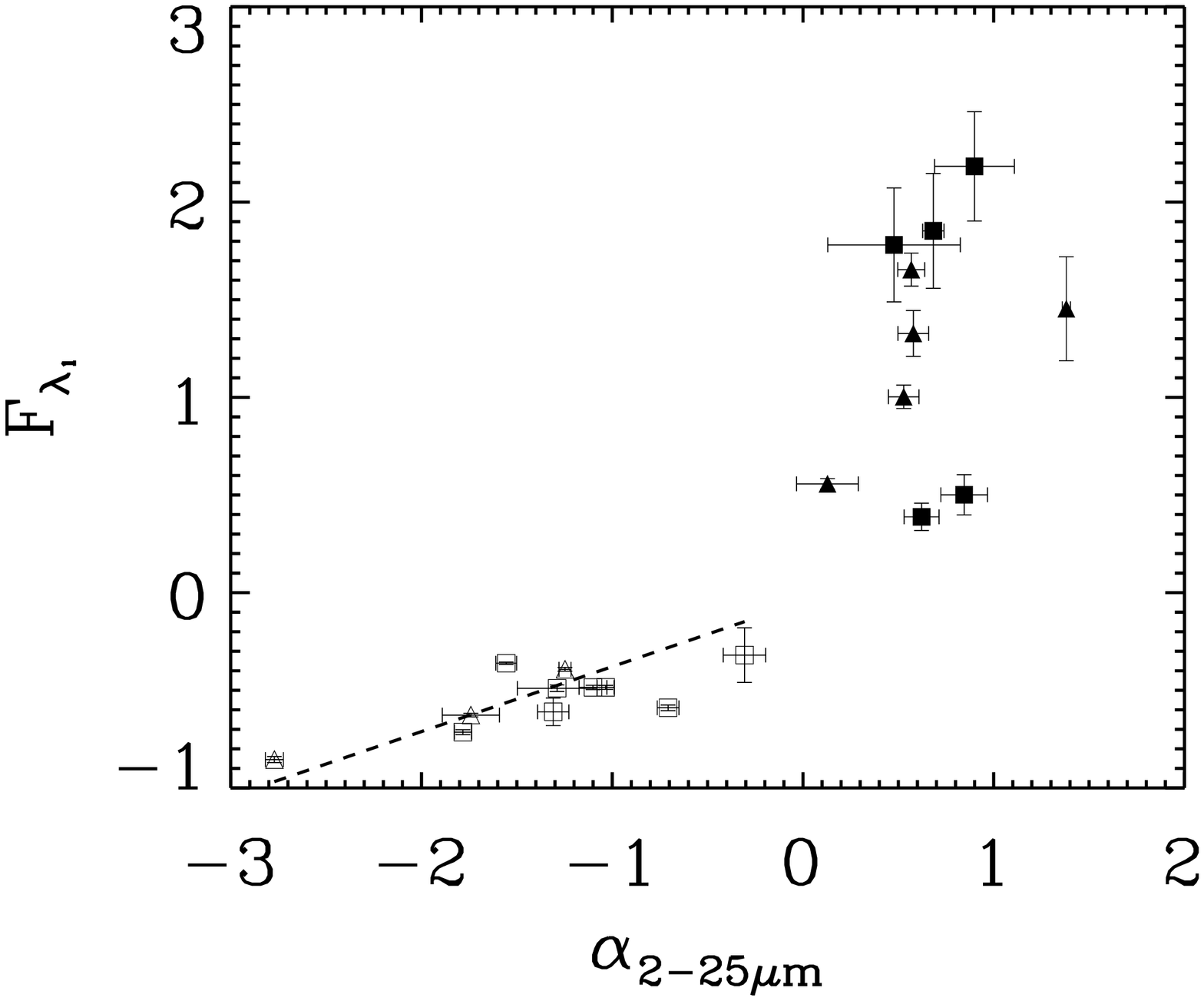}
\vspace{0.3cm}
\figcaption{Correlations with the SED.  This plot shows the correlation between the strength of the silicate feature (F$_{\lambda_1}$; defined in Table~\ref{tab:spectra}) and the 2--25 $\mu$m spectral index ($\alpha_{2-25\mu{m}}$).  The open symbols represent absorption spectra of solar-mass (squares) and intermediate-mass (triangles) stars in our sample.  The solid symbols represent emission spectra. Spectra that can be fit with both emission and absorption are omitted.  The error bars for $\alpha_{2-25{\mu}m}$ include the photometric errors as quoted in the literature.  Fluxes are calculated by averaging the continuum normalized spectrum over 0.2 $\mu$m.  The data are weighted by the inverse square of the errors shown, which include the effects of spectral S/N and the removal of the continuum.  For clarity, the error bars for GM Aur have been truncated at 1/7 of the full error (see text).
The strength of the absorption features appears to be strongly correlated with the spectral index ($r=0.75$, 1.3$\%$ probability of being drawn from random distribution), as indicated by the dashed line.  In contrast, the strength of the emission features is only weakly correlated with the spectral index, if at all; $r=0.29$, with a 40$\%$ probability that the data are drawn from a random distribution.
\label{fig:specindex}}
\end{figure}

\citet{lada_84} define a classification system using the SED slope, or 2--25 $\mu$m spectral index ($\alpha_{2-25{\mu}m}$).  The spectral index is negative for class I sources (embedded stars), positive in the range of 0--1 for class II sources (optically thick disks), and positive and large ($\sim$3) for class III sources (optically thin disks).  
In Figure~\ref{fig:specindex} we explore possible correlations between the feature strength and the spectral index for spectra with pure emission or absorption features.  Spectra showing evidence of mixed absorption and emission are not included.
The feature strengths (F$_{\lambda_1}$) used here are calculated from the normalized spectra (F$ = ($F$_\nu-$F$_c)/$F$_c$), by averaging over a region 0.2 $\mu$m wide centered on the largest absolute value ($\mid$F$_{\lambda_1}$$\mid$) in the normalized spectrum (see Table~\ref{tab:spectra}).  The spectra were smoothed before fitting to remove large fluctuations due to noise, which may have a large effect on locating the maximum.   
The quoted errors include the effects of spectral S/N (the standard deviation of the mean) as well as the effects of the continuum normalization method (the errors in determination of the endpoint fluxes).  Error bars are particularly large for GM Aur, for which the spectrum possessed low S/N near the endpoints used for continuum normalization.  For this reason, the error bars for GM Aur shown in all figures have been truncated at 1/7 of the full error. The correlation is evaluated using a ${\chi}^2$ linear least-squares regression method.
During the calculation of the regressions, each data point is weighted by the inverse square of the error in x and y, so the effects of data with large errors are minimized. One should note that this method may reduce the weighting of T Tauri star disks, as their spectra for the most part have lower S/N than do spectra of disks around HAEBE stars. 

There is a strong inverse correlation between the strength of the absorption features and the spectral index.  
As the spectral index increases, moving from class I toward the realm of class II sources, the strength of absorption features becomes shallower.
Furthermore, 4/5 of the complex spectra (absorption+emission), which are not included in Figure~\ref{fig:specindex}, possess spectral indices which are at the low end of the distribution ($\leq$0.5), suggesting that they may be in an intermediate stage between class I and class II.  This trend indicates that the strength of silicate absorption is primarily a function of 
the optical depth and not dust composition/size.
In contrast, there is no compelling correlation between strength and spectral index for the emission sources in Figure~\ref{fig:specindex}. 

Although the strength of the silicate emission features not related to the spectral index, it may still be dependent on or related to the disk morphology.
In order to determine the relationship between the strength of silicate emission and disk morphology, the ratio of the IR to stellar luminosity was calculated. L$_{IR}$/L$_{*}$ is related to the optical depth of the disk material for optically thin emission. For optically thick emission, however, L$_{IR}$/L$_{*}$ is indicative of the disk geometry, tracing the fraction of light intercepted by the disk \citep{backman_93}.  In this case, disk flaring and orientation both affect the IR-to-stellar luminosity ratio, with a maximum value of L$_{IR}$/L$_{*}$ of 1/2 for a flared disk, in which the thickness increases with distance from the star \citep{kenyon_h87}, and 1/4 for a flat optically thick disk.  
IR-to-stellar luminosity ratios for the disks in our sample, calculated as described in $\S$4, are presented in Table~\ref{tab:diskparam}.
The optically thin debris disks possess L$_{IR}$/L$_{*}$ $<$0.1 and do not show evidence of silicate emission.
We find generally higher values of L$_{IR}$/L$_{*}$ for the optically thick class II disks.  These ratios are always greater than 0.1 with a maximum of $\sim$0.5.  
The strengths of the emission features are not found to be correlated with L$_{IR}$/L$_*$.  As L$_{IR}$/L$_*$ is affected by both disk flaring and orientation, we can only say that the combined effect of the two is not clearly related to the strength of the 10 $\mu$m silicate emission feature.   

In summary, the strength of silicate absorption can be tied to the self-embeddedness of the star and is probing primarily the optical depth of the absorbing dust.
the strength of the emission features appears to trace grain properties (e.g., grain growth and/or crystallinity), as opposed to optical depth, and these properties do not appear to be directly connected to evolutionary stage, as traced by the \citet{lada_84} classification system, or to disk morphology.  This is consistent with a scenario in which the observed emission arises from an optically thin region on disk surfaces.

\subsection{Width and Central Wavelength of the Silicate Feature}

For 6 lines of sight in the ISM, \citet{bowey_02} find that the FWHM of the observed silicate absorption decreases as the features shift to longer wavelength. They also find that these parameters are strongly related to the stellar environment. Three spectra probing the diffuse medium showed silicate 
absorption features with minima at $\sim$9.8 $\mu$m and FWHM of $\sim$2.5 $\mu$m, typical of amorphous olivine. The others, within the Taurus molecular cloud, possess silicate absorption features near 9.6 $\mu$m with FWHM of 3.3 $\mu$m, 
typical of amorphous pyroxene.  
The authors suggest that this trend represents an evolution from olivine-dominated silicates in the diffuse ISM to pyroxene-dominated silicates in circumstellar environments.  

In order to search for trends in the silicate composition as a function of {\it circumstellar} environment, we perform similar calculations for the silicate features observed toward our sample of 17 embedded YSOs and 10 optically thick disks. Sources with photospheric spectra are not included. 
As discussed above, the strength (and shape) of silicate absorption features is a function of the optical depth.  The central wavelength of the absorption, however, is dependent on the dust composition.  
In contrast, the strength and width of silicate emission features (if optically thin) is almost solely dependent on the dust cross section, which is affected by both grain composition and size.  The peak of the emission is also believed to be a function of both grain size and composition \citep{vanboekel_03}.  

The silicate feature in each normalized spectrum (F$ = ($F$_\nu-$F$_c)/$F$_c$) was fit by a gaussian to find the FWHM, central wavelength ($\lambda_{2}$), and the normalized flux (F$_{\lambda_2}$) corresponding to that central wavelength.  To account for the structured nature of the silicate features observed, we also record the wavelength ($\lambda_{1}$) corresponding to the largest absolute value ($\mid$F$_{\lambda_1}$$\mid$; used in $\S$6.1) in the normalized spectrum. This provides a better characterization in the case of dust with large crystalline-to-amorphous silicate emission ratios (e.g., for HD 179218) where F$_{\lambda_2}$ lies near the center of the feature at $\lambda_{2}\sim$10.5 $\mu$m, but F$_{\lambda_1}$ successfully traces the stronger flux near the forsterite peak ($\lambda_{1}\sim$11 $\mu$m).  
The results of these calculations are shown in Table~\ref{tab:spectra}.

\begin{deluxetable}{lcccccc}
\tablecaption{Silicate spectral parameters \label{tab:spectra}}
\tablehead{\colhead{}  & \colhead{$\lambda_{1}$} & \colhead{$\lambda_{2}$} & \colhead{} & \colhead{} & \colhead{} & \colhead{} \\
\colhead{Source}  & \colhead{($\mu$m)} & \colhead{($\mu$m)} & \colhead{F$_{\lambda_1}$} & \colhead{F$_{\lambda_2}$} & \colhead{FWHM} & \colhead{Fit}} 
\startdata
IRAS 04016+2610  & 9.34  & 9.68  &-0.49  &-0.47  &2.43  &abs      \\ 
IRAS 04108+2803B\tablenotemark{a} & 9.81  &9.77   &-0.13  &-0.15  &0.36  &abs      \\
		 &11.67  &11.44  & 0.16  & 0.15  &1.34  &em    \\
IRAS 04169+2702  & 9.60  & 9.58  &-0.49  &-0.46  &2.00  &abs      \\
IRAS 04181+2654A & 9.18  & 9.52  &-0.32  &-0.26  &2.26  &abs      \\
IRAS 04181+2654B\tablenotemark{a} & 9.26  &9.28   &-0.31  &-0.30  &0.93  &abs      \\
		 &11.67  &11.44  & 0.16  & 0.10  &1.34  &em    \\
IRAS 04239+2436  & 9.22  & 9.72  &-0.48  &-0.45  &2.48  &abs      \\
IRAS 04248+2612\tablenotemark{a}  & 8.74  &8.82   &-0.19  &-0.16  &0.85  &abs      \\
		 &10.00  &10.43  & 0.22  & 0.20  &1.34  &em    \\
IRAS 04264+2433  &10.26  &10.07  & 0.54  & 0.51  &2.70  &em    \\
Haro 6-10A       & 9.41  & 9.75  &-0.59  &-0.59  &2.36  &abs      \\
Haro 6-10B       & 9.37  & 9.64  &-0.36  &-0.35  &2.20  &abs      \\
IRAS 04287+1801  & 9.23  & 9.93  &-0.71  &-0.69  &3.17  &abs      \\
IRAS 04295+2251\tablenotemark{a}  & 9.37  &9.20   &-0.14  &-0.14  &0.72  &abs      \\
		 &10.63  &10.78  & 0.17  & 0.18  &1.26  &em    \\
AA Tau           &10.15  &10.09  & 0.50  & 0.49  &2.54  &em    \\
LkCa 15          & 9.96  &10.19  & 2.18  & 2.22  &2.20  &em    \\
IRAS 04381+2540  & 9.71  & 9.99  &-0.61  &-0.57  &2.93  &abs      \\
IRAS 04489+3042\tablenotemark{a}  & 9.40  &9.20   &-0.14  &-0.15  &0.90  &abs      \\
		 &11.69  &11.52  & 0.19  & 0.23  &2.81  &em    \\
GM Aur           & 9.77  & 9.87  & 1.78  & 1.68  &2.19  &em    \\
MWC 480          &10.58  &10.32  & 1.00  & 1.02  &2.45  &em    \\
BN               & 9.63  & 9.69  &-0.39  &-0.41  &1.76  &abs      \\
NGC 2024 IRS2    & 9.67  & 9.87  &-0.63  &-0.63  &2.35  &abs      \\
Mon R2 IRS3      & 9.82  & 9.93  &-0.85  &-0.90  &2.93  &abs      \\
Hen 3-600A       &10.98  &10.62  & 0.39  & 0.32  &2.07  &em    \\
IRAS 14050-4109  & 9.62  & 9.95  & 1.85  & 1.86  &2.15  &em    \\
HD 163296        & 9.61  &10.02  & 1.33  & 1.34  &2.48  &em    \\
HD 179218        &11.01  &10.48  & 0.56  & 0.53  &2.39  &em    \\
WW Vul           & 9.84  & 9.92  & 1.65  & 1.63  &2.33  &em    \\
HD 184761        &10.06  &10.07  & 1.45  & 1.46  &2.20  &em    \\
\enddata  
\tablenotetext{a}{The observed spectrum can be fit by absorption + emission.}
\tablecomments{$\lambda_{1}$ and F$_{1}$ indicate central wavelengths and feature strengths calculated by fitting a gaussian to the normalized spectrum. $\lambda_{2}$ and F$_{2}$ indicate central wavelengths and feature strengths calculated by finding the largest absolute value of the normalized flux. In all cases, FWHM are found from gaussian fits.}
\end{deluxetable}

This analysis is not as straight-forward for the spectra that show evidence of {\it both} emission and absorption. If we fit the absorption and emission components individually for the complex spectra, as shown in Table~\ref{tab:spectra}, we find small FWHM for the individual features, which are shifted to shorter/longer wavelengths for absorption/emission.   
For all 5 of the complex spectra we fit absorption components with  $\lambda_{1,2}<$ 10 $\mu$m and emission components with $\lambda_{1,2}\gtrsim$ 10 $\mu$m.
Although these complex spectra represent an interesting new class of sources, the emission and absorption features clearly can not be deconvolved empirically.
Source geometry must be taken into account in the interpretation of these spectra.
For these reasons, we include only pure absorption and emission features in the rest of our analyses.

\begin{figure}[t]
\includegraphics[angle=90,width=9.5cm]{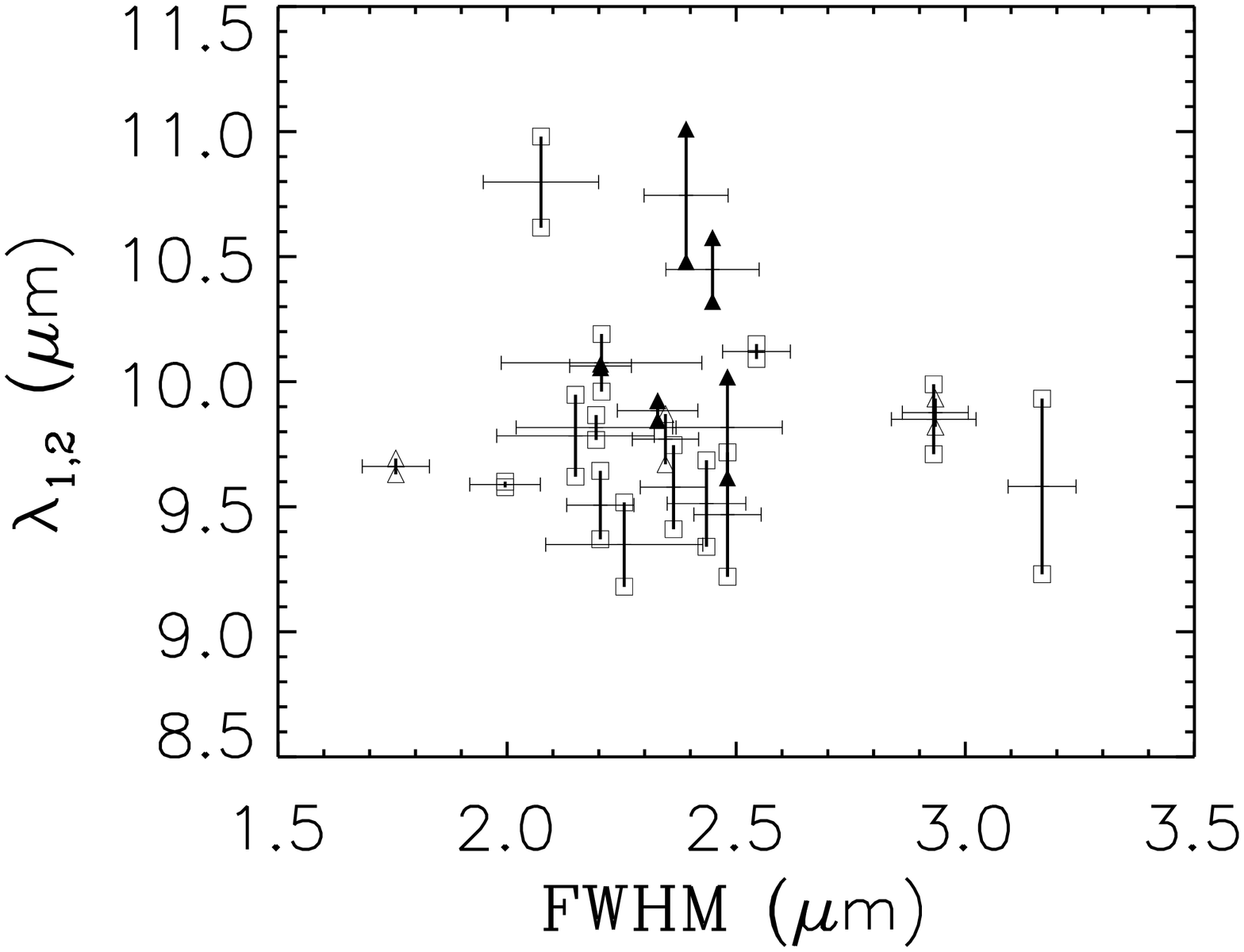}
\vspace{0.3cm}
\figcaption{Plot of the wavelength of the emission/absorption maxima ($\lambda_{1,2}$) and the width of the feature (FWHM), as presented in Table~\ref{tab:spectra}. $\lambda_{1}$ and $\lambda_{2}$ are represented by connected pairs. Symbols are as described in Figure~\ref{fig:specindex}.  Neither the sample of silicate absorption spectra nor that of silicate emission spectra show significant correlation between $\lambda_{1,2}$ and FWHM.  The two samples do show different trends, however, with the emission spectra spanning a larger range in central wavelength and the absorption spectra spanning a larger range in FWHM.
\label{fig:bowey}}
\end{figure}

In Figure~\ref{fig:bowey}, the central wavelengths ($\lambda_1$ and $\lambda_2$) of the emission/absorption maxima (as described by F$_{\lambda_1}$ and F$_{\lambda_2}$) are plotted with respect to the FWHM. 
Although a subset of the pure absorption or emission features are clustered about 9.75$\pm$0.15 $\mu$m, the sample encompasses a large range of central wavelength ($\sim$9-11 $\mu$m) and FWHM ($\sim$1.7-3 $\mu$m), suggesting a range of source environments. 
There appears to be no strong correlation between $\lambda_{1,2}$ and FWHM in the sample as a whole ($r=0.04$), nor in either sub-population of silicate absorption and emission spectra. 
The two samples do show different trends, however, with the emission spectra spanning a larger range of central wavelength and the absorption spectra spanning a larger range of FWHM.  To address the nature of these trends in more detail, we now discuss the absorption and emission subsamples in turn.

The silicate absorption features show little variation in central wavelength (9.6$\pm$0.2 $\mu$m), which is consistent with amorphous olivines and pyroxenes.  The trend seen by \citet{bowey_02} is not seen, suggesting that there is no clear correlation within this sample of class I objects between the grain composition (traced by $\lambda_{1,2}$) and optical depth (traced by FWHM).
The FWHM of the absorption features varies significantly ($\sim$1.8--3.2 $\mu$m), indicating a range of optical depths, indicative of varying envelope thickness or morphology.
There are three outlying absorption sources possessing large FWHM (IRAS 04287+1801, IRAS 04381+2540, and Mon R2 IRS3). These were previously identified in $\S$5.1 as possessing additional shallow feature near 11.2 $\mu$m and the very large FWHM may be merely a result of the blending of the 11.2 $\mu$m ice feature with the silicate feature.

The silicate emission features in general peak at longer central wavelength (10.2$\pm$0.3 $\mu$m) in comparison to the embedded sources, indicating either grain growth or modified composition.
The silicate emission features also show larger variations in the central wavelength, but little variation in the FWHM.
The FWHM for the emission features is largely dependent on the grain cross section and, thus, on both the grain size and composition.  The range in FWHM of roughly 2 to 2.5 $\mu$m is consistent with primarily amorphous silicate grains of 0.1--2 $\mu$m sizes \citep{vanboekel_03}.
The variations in FWHM are also consistent with those expected from the addition of crystalline silicates.  Indeed, the 2 emission spectra with the largest peak wavelengths ($\sim$ 11.2 $\mu$m) are those previously identified as possessing strong crystalline silicate emission, Hen 3-600A and HD 179218.  The emission spectrum from MWC 480 is also an outlier, peaking at 10.3--10.5 $\mu$m, which is consistent with crystalline enstatite.  

Thus, sources with silicate absorption features have central wavelengths and FWHM  generally similar to amorphous olivine or pyroxene, while those with silicate emission show increased peak wavelengths and variations in FWHM characteristic of more processed dust.
The relationship between the silicate emission feature and size and compositional variations of the dust will be explored in more detail below.
We will not discuss the absorption or the absorption-plus-emission sources any further.

\subsection{Spectral Color-Color Diagrams:  Grain Composition}

\begin{figure}
\includegraphics[angle=90,width=9.5cm]{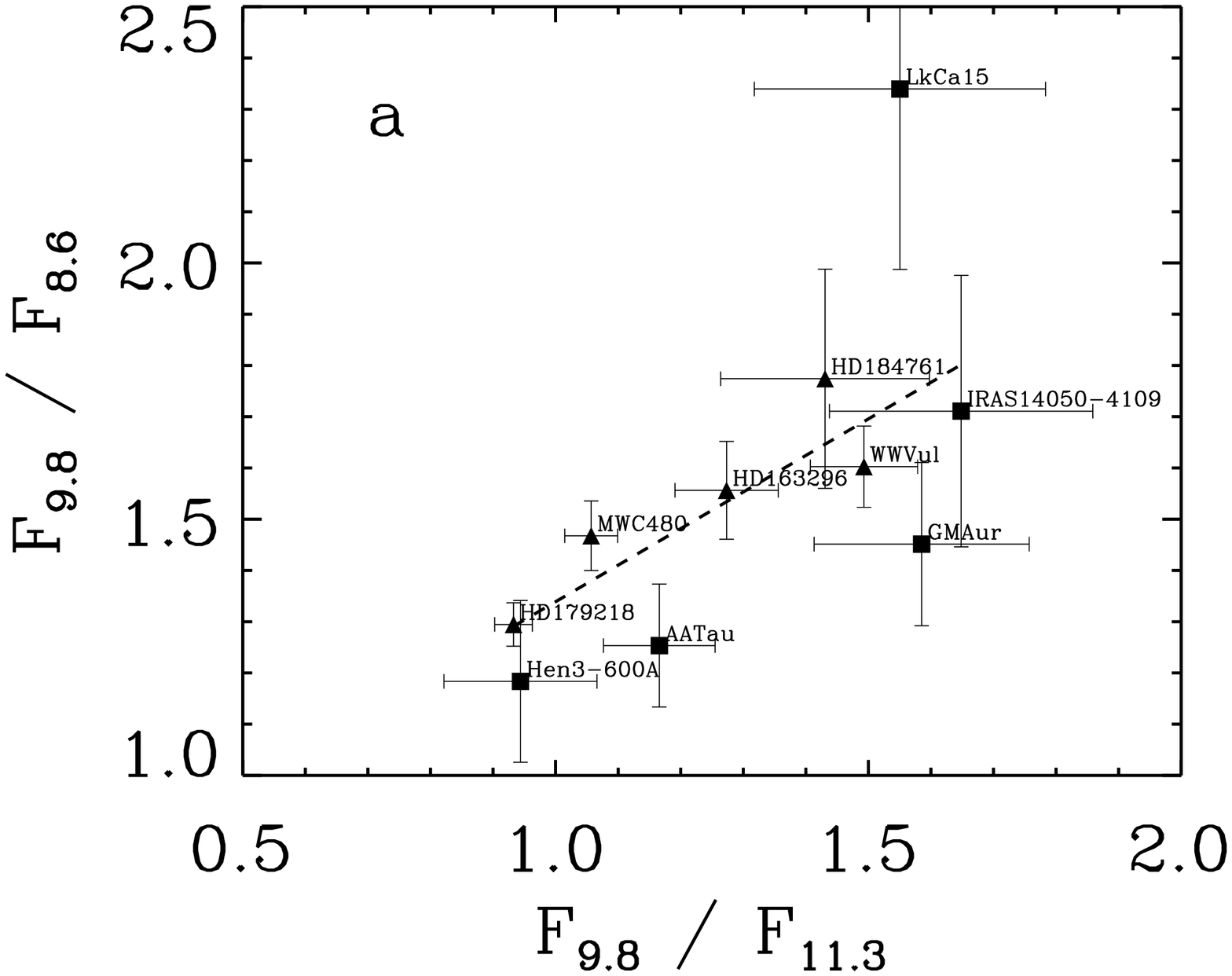}
\vspace{0.3cm}
\includegraphics[angle=90,width=9.5cm]{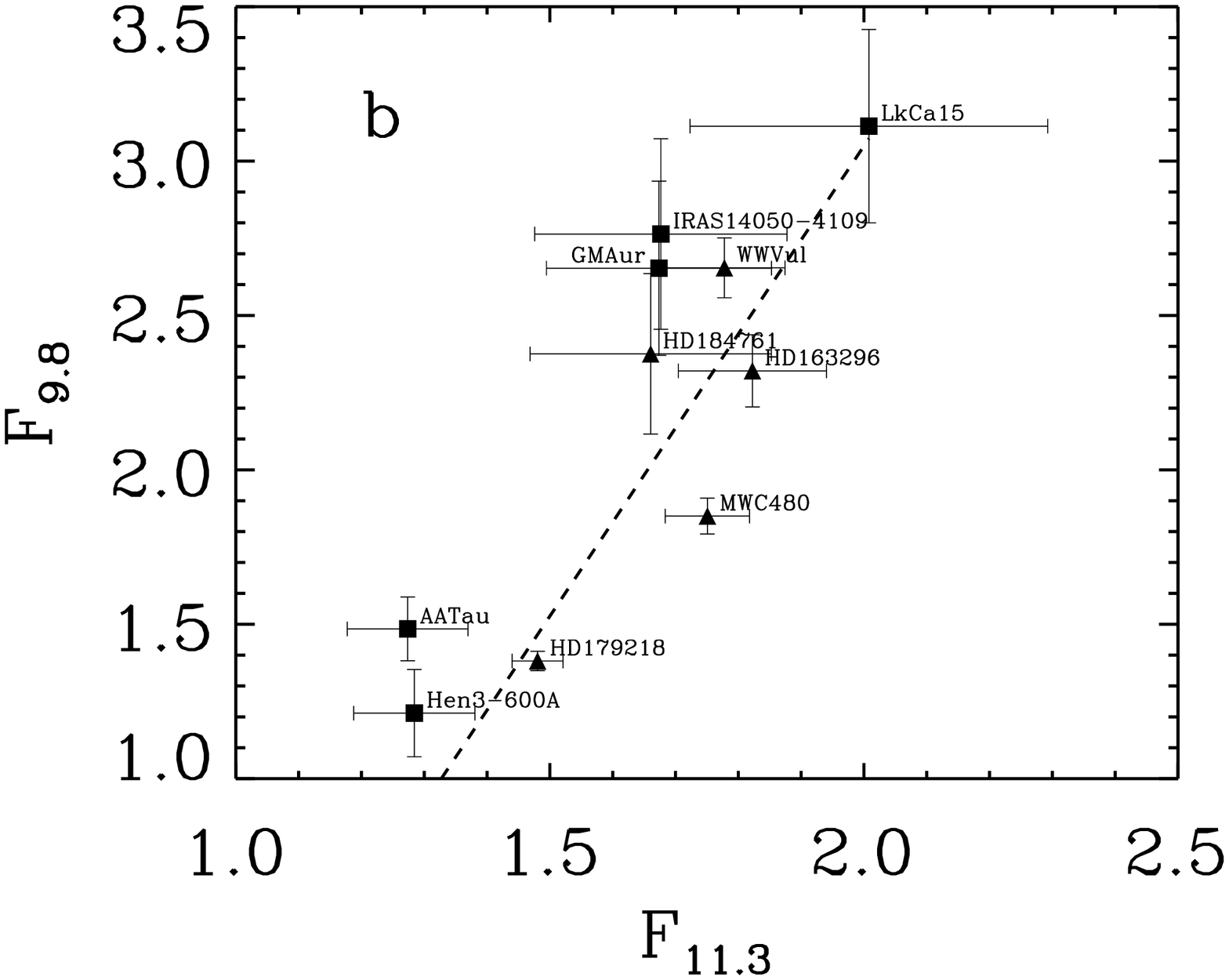}
\figcaption{IR color-color diagrams for silicate emission.  (a) The relationship between the slopes on the long (F$_{9.8}$/F$_{11.3}$) and short wavelength (F$_{9.8}$/F$_{8.6}$) side of the normalized spectra. A linear least-squares fit to the data is depicted with a dashed line, (F$_{9.8}$/F$_{8.6}$) = (0.7$\pm$0.2)$\times$(F$_{9.8}$/F$_{11.3}$) + (0.6$\pm$0.2), and characterized by a correlation coefficient of $r=0.67$. 
 (b) Fluxes corresponding to amorphous olivine (F$_{9.8}$) and crystalline forsterite (F$_{11.3}$) in the normalized spectra.  The linear fit to these data (dashed line, $r=0.85$), F$_{9.8}$ = (3.0$\pm$0.6) $\times$ F$_{11.3}$ - (3.0$\pm$1.0), indicates that F$_{11.3}$ does not increase at the expense of F$_{9.8}$.  Fluxes and corresponding error bars are calculated as in Figure~\ref{fig:specindex}.
\label{fig:colorcorr}}
\end{figure}

Now we attempt to use fluxes derived from the continuum normalized silicate emission features as probes of grain composition, following the methods used for ISO spectra of HAEBE stars \citep{bouwman_01}.  
In particular, we will compare fluxes at the nominal wavelengths of emission from amorphous and crystalline olivine, 9.8 and 11.3 $\mu$m, and ratios of 9.8-to-8.6 versus 9.8-to-11.3 $\mu$m fluxes in an attempt to assess the probe the crystallization mechanism.
 Figure~\ref{fig:colorcorr} displays flux and color-color plots for the silicate emission features observed toward our sample.  
The fluxes are calculated by averaging over 0.2 $\mu$m. The errors are calculated as described in $\S$6.1 and include the affects of S/N and the continuum subtraction method.

In Figure~\ref{fig:colorcorr}a, the relationship between the blue and red slopes is examined.  Assuming the emission arises primarily from amorphous olivine (peaking near 9.8 $\mu$m), the slopes on the long and short wavelength side of the normalized spectra can be approximated by F$_{9.8}/$F$_{11.3}$ and F$_{9.8}/$F$_{8.6}$, respectively. The dashed line indicates the correlation ($r=0.68$) between the blue and red slopes for the observed silicate emission features.  
This trend implies that decreasing F$_{9.8}/$F$_{8.6}$ (blue slopes) correspond to decreasing F$_{9.8}/$F$_{11.3}$ (red slopes), indicating that the entire feature becomes flatter as emission at 11.3 $\mu$m becomes more prominent. 
The two sources with silicate features that peak above 11 $\mu$m, HD 179218 and Hen 3-600A, also fall along this trend, occupying the lower left corner of the plot.  Although the variations in the structure of the silicate emission has often been attributed to variations in crystalline fraction, this trend is consistent with grain compositions dominated by amorphous olivine, with variations due to changes in grain size \citep[e.g.,][]{vanboekel_03}.  

If the above trend can be attributed to the conversion from dust dominated by amorphous olivine to that dominated by crystalline forsterite, then variations in the crystallinity of the dust in these sources should be evident in the relationship between the fluxes at these wavelengths.  The fluxes corresponding to the peaks of the amorphous olivine (F$_{9.8}$) and crystalline forsterite features (F$_{11.3}$) for the normalized spectra are shown in Figure~\ref{fig:colorcorr}b. 
There is a linear correlation between the fluxes at 11.3 and 9.8 $\mu$m indicating that the flux at 11.3 $\mu$m does {\it not} increase at the expense of the flux at 9.8 $\mu$m, as we would expect from crystallization of amorphous olivines (emitting near 9.8 $\mu$m).  
Additionally, the spectra with dominant emission at 11.3 $\mu$m, are weaker overall and fall in the lower left corner of the plot.  
This suggests that the crystalline silicates observed toward these sources either are not produced from the same reservoir of small amorphous grains that emits at 9.8 $\mu$m, or that variation in the shape of the silicate feature is not dominated by crystallization.  

The above analysis demonstrates that the 9.8-to-8.6 and 9.8-to-11.3 colors are correlated within the observed silicate emission features. The features become flatter as the relative emission at 11.3 $\mu$m increases, which is consistent with increased grain sizes. Furthermore, the flux at 11.3 $\mu$m does not decrease at the expense of that at 9.8 $\mu$m, indicating that if the flattening of the feature is due to crystallization, then the crystalline forsterite emitting at 11.3 $\mu$m are not crystallized from the amorphous silicate population emitting at 9.8 $\mu$m.

\subsection{Strength and Shape: Grain Size}

\begin{figure}[h]
\includegraphics[angle=90,width=9.5cm]{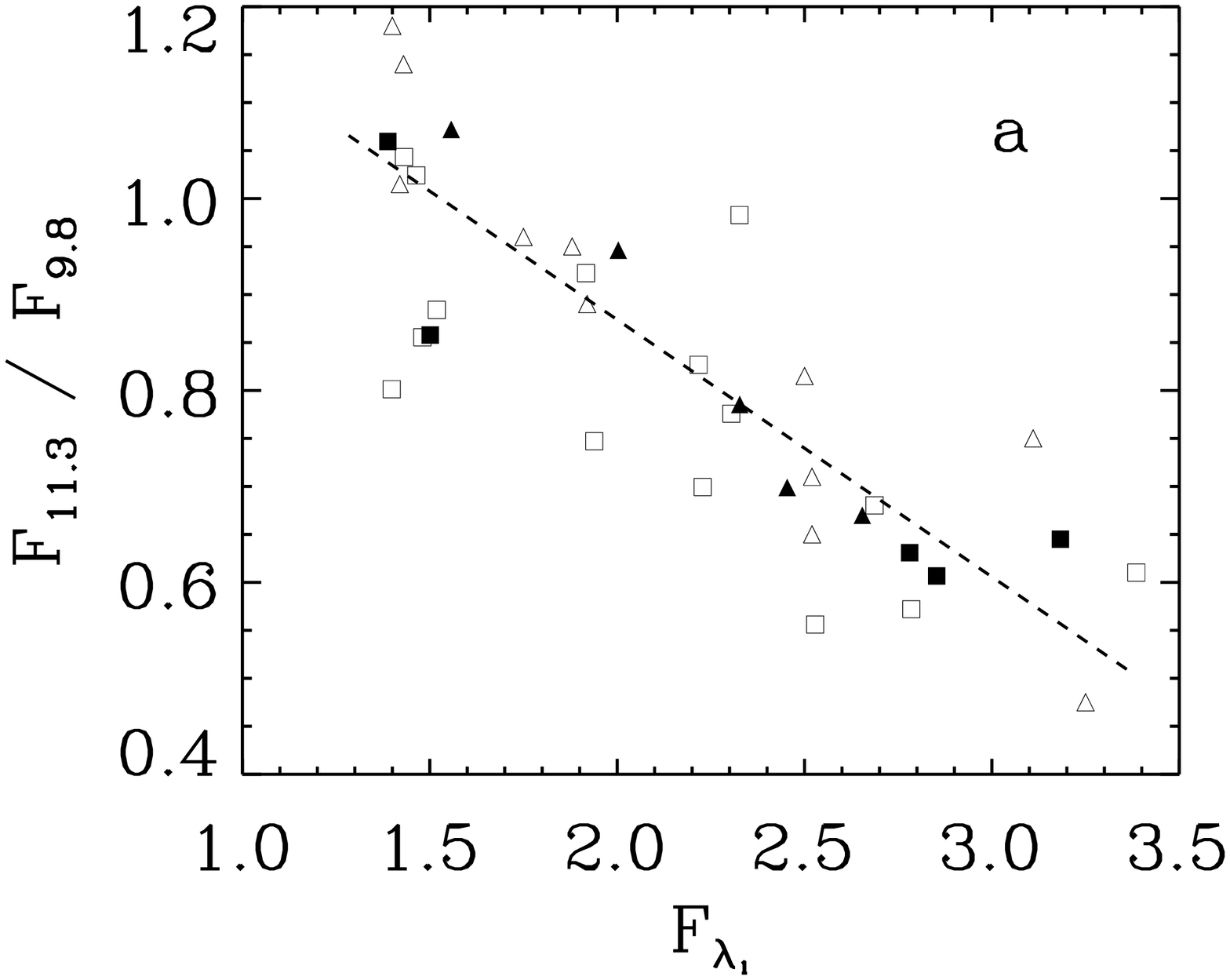}
\includegraphics[angle=90,width=9.5cm]{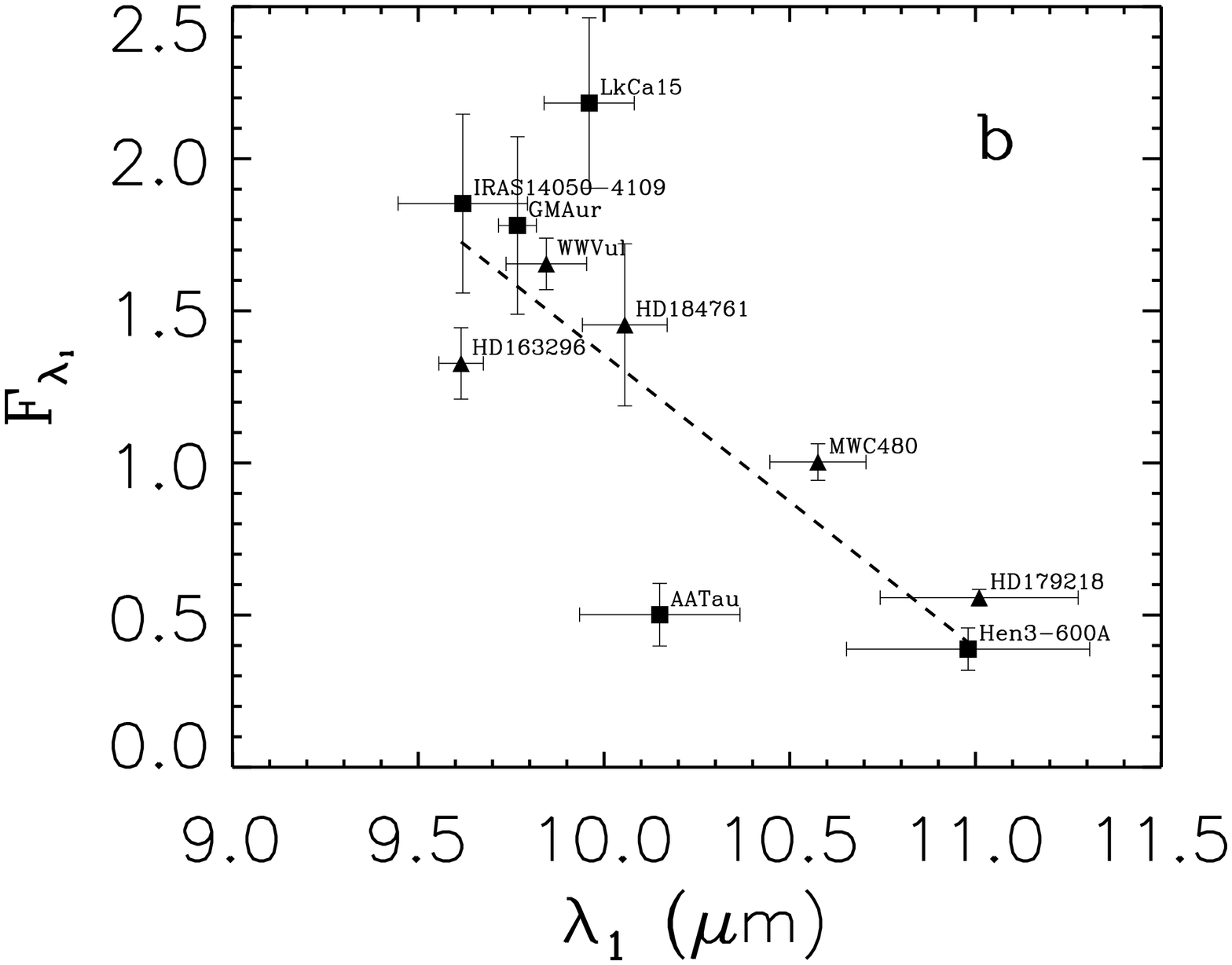}
\vspace{0.3cm}
\figcaption{Emission feature characteristics. In both plots, the solid squares and triangles  represent the observations of T Tauri and HAEBE stars in our sample.  (a) The correlation between feature shape (F$_{11.3}/$F$_{9.8}$) and strength (F$_{\lambda_1}$), as noted previously for HAEBE \citep[open triangles]{vanboekel_03} and T Tauri stars \citep[open squares]{przygodda_03}. The dashed line is a linear fit to the entire sample, F$_{11.3}/$F$_{9.8}$=(1.41$\pm$0.07)$\times$F$_{\lambda_1}$-(0.27$\pm$0.03), with a correlation coefficient of $r=0.82$.  (b) The strength of the silicate feature decreases with decreasing peak wavelength ($\lambda_{1}$) as, F$_{\lambda_1}$ = (-1.0$\pm0.1)\times\lambda_{1}$ + (11$\pm$1), with $r=0.78$. Both trends are consistent with grain growth (see text).  Fluxes and corresponding error bars are calculated as in Figure~\ref{fig:specindex}.
\label{fig:vbfit}}
\end{figure}

Here we discuss the possibility that the changes in the silicate emission feature shape discussed above are due to changes in grain size, rather than crystallinity.
Although observations of broadened 8--13 $\mu$m silicate features with emission observed at 11.3 $\mu$m were long believed to be an clear indication of the presence of crystalline forsterite, the analysis presented in $\S$6.3 and recent studies by \citep{vanboekel_03} suggest that such features are actually representative of increased grain size.
These studies show that the mid-IR absorption coefficient of $\sim$0.1 $\mu$m-sized amorphous olivine grains possesses a triangular shape similar to that observed in the diffuse ISM.  
Due to a general decrease in the absorption cross section, a population of larger amorphous olivine grains ($\sim$2.0 $\mu$m) has an absorption coefficient that is weaker by a factor of $\sim$2.3 and flattened between 9.5 and 12 $\mu$m.  
Examination of the emission spectra of 12 HAEBE stars showed that there is a correlation between the feature strength (peak flux of the normalized spectrum) and shape (as indicated by the ratio of the flux at 11.2 to 8.9 $\mu$m), with broader emission features being weaker as shown in Figure~\ref{fig:vbfit}a (open triangles).  
\citet{przygodda_03} found a similar trend for a sample of disks surrounding low-mass stars (open squares). 

We perform a similar analysis for our sample of T Tauri and HAEBE stars, which are shown as solid squares and triangles in Figure~\ref{fig:vbfit}a.
Fluxes and errors are calculated as described in $\S$6.3.
The entire sample of both T Tauri and HAEBE star spectra can be fit by a single trend.
This trend is consistent with variations in grain size among the sample, but we can not rule out the possibility that the changes in feature shape and strength result from an increased crystallization fraction ($\S$6.3).
The two sources with spectra consistent with forsterite emission at 11.3 $\mu$m do not deviate significantly from this trend.  
It is therefore reasonable to think of the strong, narrow and the broad, weak silicate emission features as arising from ``unprocessed'' and ``processed'' dust, respectively.
The contributions from dust size and composition can thus only be separated when emission from crystalline silicates are clearly seen in the spectrum as an isolated feature at 11.3 $\mu$m.

Although generally used as a measure of the crystallization fraction in the emitting silicate dust (e.g., the ``9.8-to-11.3 phase transition''), the wavelength of maximum emission may also be affected by grain size.  
In Figure~\ref{fig:vbfit}b the maximum emission in the normalized spectrum F$_{\lambda_1}$ is plotted with respect to the wavelength corresponding to the maximum $\lambda_{1}$.
Figure~\ref{fig:vbfit}b indicates that for our sample as the feature maximum shifts to longer wavelengths, the strength of the emission feature decreases.  
This trend is consistent with the effects of grain growth. As the grain size increases, the feature flattens and the maximum decreases in flux and shifts to longer wavelength.
Although, the two sources with strong crystalline forsterite features (HD 179218 and Hen 3-600A) have weaker peak flux and show up at the bottom right in Figure~\ref{fig:vbfit}b, 
in general moving along the trend from the upper left to lower right of Figure~\ref{fig:vbfit}b does not coincide with the appearance of emission from crystalline forsterite.
HD 163296, which has a visible bump near 11.3 $\mu$m, has a silicate feature which is still quite strong (in the upper left of the plot).  Furthermore, the broad silicate emission feature observed toward AA Tau, which shows no evidence of emission from crystalline forsterite, is quite weak and lies in the middle of the trend.  
 Therefore, the shift in the wavelength of maximum emission in the 10 $\mu$m region, or the ``9.8-to-11.3 $\mu$m phase transition'', appears to be largely dependent grain size variations and thus is in fact {\it not} a  good indicator of crystallization fraction.

If the trends depicted in Figure~\ref{fig:vbfit} primarily indicate a change in the grain size, then they suggest a progressive removal of small grains ($<$ 2 $\mu$m) from the warm (T $\approx$ 300K) disk surface probed by the 10 $\mu$m  feature.  
Dust settling would preferentially remove larger grains from the disk surface. 
The absence of small grains could be related to the removal of dust through gap clearing, but if so we would expect that GM Aur, which shows evidence of a substantial gap, to display a broadened feature indicative of larger grains.  The flat, broadened spectra in the bottom right of Figure~\ref{fig:vbfit}b are therefore most likely a result of grain growth in the surfaces of these disks.

\section{CONCLUSIONS AND FURTHER WORK}

8--13 $\mu$m spectra were obtained for 34 low- and intermediate-mass YSOs.  
The evolutionary sequence seen in ISO observations of high/intermediate-mass YSOs \citep{meeus_99}, was confirmed and extended here to $\sim$solar-mass stars.  Amorphous silicate absorption is observed toward young, embedded sources. In 5 of 17 embedded sources, complex spectra that may be interpreted as a combination of silicate absorption and emission are observed.  These spectra may represent an interesting new class of sources that are in between Class I and Class II, but require physical models to deconvolve the emission and absorption.
Silicates are observed in emission toward optically revealed stars.  
The shape of these emission features varies from strong triangular emission peaked near 9.8 $\mu$m, to flat, broad emission with a central wavelength between 10 and 11 $\mu$m.  Finally, silicate emission was absent for the debris disks in our sample.

To aid in the interpretation of the silicate features, we explore the connections between silicate feature properties with stellar evolution through comparison with the spectral energy distributions.
We find that the strength of 10 $\mu$m silicate absorption is strongly correlated with the 2--25 $\mu$m spectral index.
Class I sources with complex absorption+emission features have small spectral indices when compared to those with pure absorption, suggesting that the former are less embedded.
These results indicate that the silicate absorption feature is indeed a good probe of the circumstellar environment, tracing the self-embeddedness of the star.
In contrast, the strength of the silicate emission feature does not appear to be correlated with the 2--25 $\mu$m spectral index (evolutionary stage) or the IR-to-stellar luminosity ratio (disk geometry) and thus probes grain properties rather than source morphology.

Although the strength of the silicate absorption features is dependent on the circumstellar environment, the central wavelength of the emission should be sensitive to the grain composition.  
We therefore attempted to quantify the effects of grain properties, such as size and composition, on the shape and/or central wavelength of all of the silicate features observed in our sample.
In general, the features are clustered around 9.7 $\mu$m, 
corresponding to amorphous olivine, with FWHM near $\sim$2.3 $\mu$m.  
A direct correlation between FWHM and central wavelength is seen for neither silicate emission nor absorption features.  
The pure silicate absorption features exhibit a large range of FWHM, likely related to differences in optical depth, and narrow range of central wavelengths, indicating limited variation in grain size and/or composition from that of the ISM.  
A larger scatter and higher mean central wavelength is found for the emission spectra, likely related to compositional variations within the sample (i.e., fraction of amorphous vs. crystalline silicates), indicating increased dust processing.

As the emission features are more sensitive to grain properties, we further analyzed the structure of the emission features, finding:

\renewcommand{\labelitemi}{---}
\begin{itemize}
\item The shape of the silicate emission of most T Tauri stars was similar to that previously observed in the diffuse ISM, in contrast to the structured emission feature observed for most HAEBE stars. This suggests an absence of small ($<10~\mu$m), warm (T$\approx$300 K) {\it crystalline} silicates in these T Tauri star disks, but colder or larger crystalline silicate grains  may still be present and, if so, will be visible through far-infrared lattice modes accessible from space born observations.  

\item A correlation between the peak shape as traced by F$_{11.3}/$F$_{9.8}$ and feature strength for 5 T Tauri and 5 HAEBE in our sample was also found. This is consistent with that seen for several HAEBE stars \citep{vanboekel_03} and T Tauri stars \citep{przygodda_03}, which are attributed to changes in grain size (from 0.1--2 $\mu$m). The entire sample can be fit best with a single trend, including spectra with strong indications of crystalline forsterite.  The wavelength of maximum emission was also found to be correlated with peak strength for our sample.  This indicates that interpretation of the  ``9.8-to-11.3 $\mu$m phase transition'' must therefore include analysis of grain size in addition to crystallization fraction.

\item We also demonstrate that the flux at 11.3 $\mu$m does not increase at the expense of that at 9.8 $\mu$m. If crystallization is dominating the transition, then this indicates that the crystalline forsterite emitting near 11.3 $\mu$m was likely not formed from the population of amorphous olivine emitting near 9.8 $\mu$m, suggesting a more complex crystallization mechanism.
Another interpretation is that the relationship between the 11.3 and 9.8 $\mu$m fluxes is {\it dominated} by grain size effects and not crystallization.
\end{itemize}

Silicate emission features in solar-mass T Tauri stars have primarily been observed  in the 10 $\mu$m atmospheric window (as in this study).
Due to sensitivity limitations, ISO studies of silicates 
focused on intermediate or high-mass stars.  The Spitzer Space Telescope will be used to expand these studies for a large sample of low-mass stars, ranging from embedded protostars to optically thin disks, via the combination of various GTO programs and the C2D ($\tau_{dust}>1$) and FEPS ($\tau_{dust}<1$) Legacy programs\footnotemark.
\footnotetext{Details of the C2D and FEPS programs can be found at http://www.peggysue.as.utexas.edu/SIRTF and http://www.feps.as.arizona.edu, respectively.} 
In this manner, the study of  the crystalline and amorphous silicates in disks, will be expanded to include disks around low-mass, sun-like stars, creating a database analogous to ISO studies of  high/intermediate-mass stars.  These studies will include longer wavelength silicate bands, which depend intricately on the coordination of the silicon atoms and can provide more specific information about the minerals present, including the Mg/Fe ratio. The mineral content of the grains in each stage of evolution can thus be directly compared to that of meteorites, asteroids and comets.

\acknowledgements
Support for JEK-S was provided by NASA through the Graduate Student Researchers Program, NGT5-50231, and through the Spitzer Space Telescope Postdoctoral Fellowship Program, under award 1256316.  GAB acknowledges additional support from the NASA Origins program. MRM acknowledges support from NASA contract 1224768 to the University of Arizona administered through the Jet Propulsion Laboratory.
The data presented herein were obtained at the W.M. Keck Observatory, which is operated as a scientific partnership among the California Institute of Technology, the University of California and NASA.  The Observatory was made possible by the generous financial support of the W.M. Keck Foundation.  The authors would also like to thank Randy Campbell for his assistance with Keck LWS operation and data reduction. The authors additionally acknowledge Jonathan Foster for his assistance with the 10 and 20 $\mu$m photometric reduction and Josh Eisner for access to his SED database of Class I sources.  The authors would also like to thank the reviewer for many useful comments.



\appendix

\section{Variability of 8--13 $\mu$m Silicate Emission Spectra}

\begin{figure*}[ht]
\hspace{1.5cm}
\includegraphics[angle=90,width=8cm]{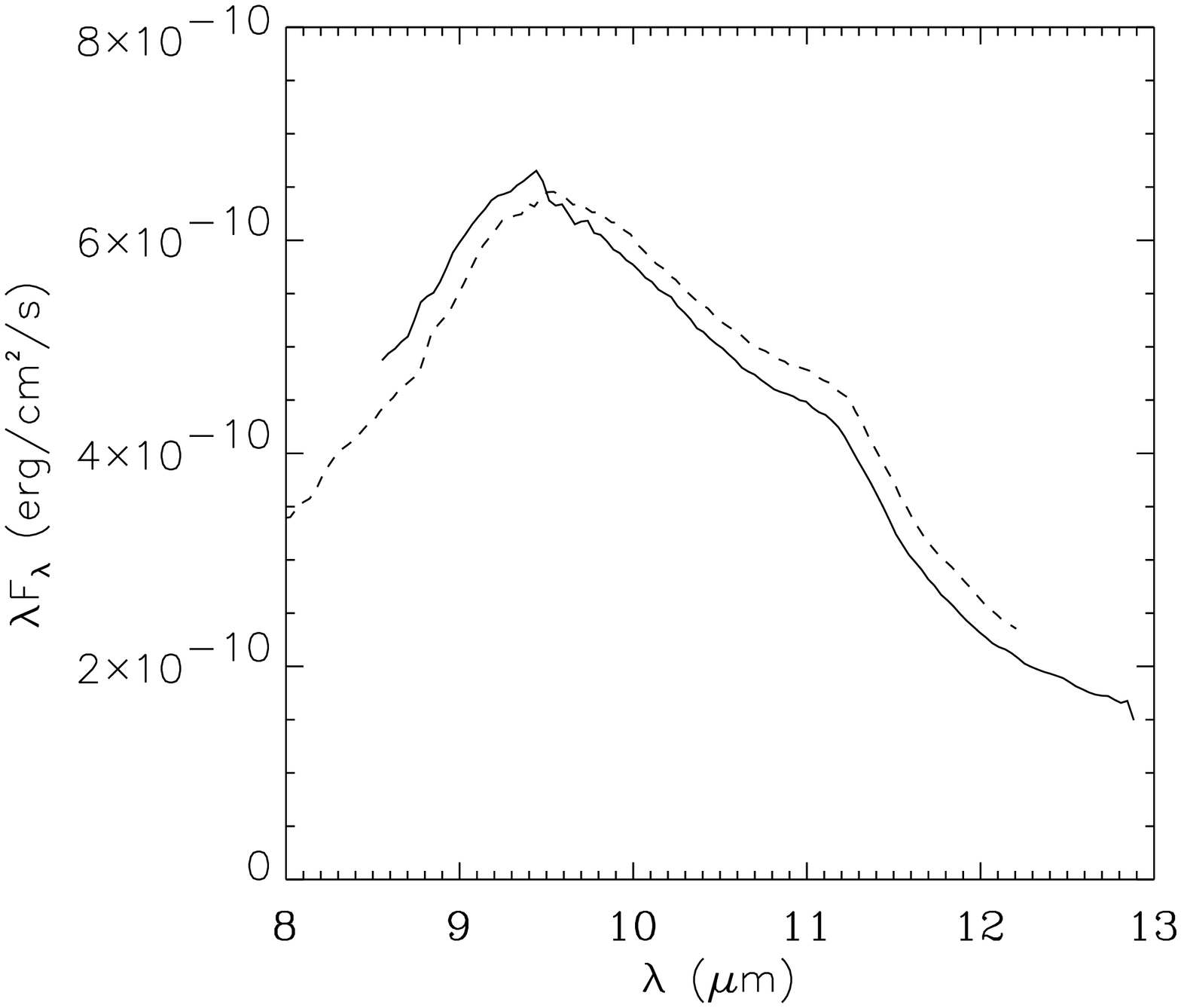}  
\includegraphics[angle=90,width=8cm]{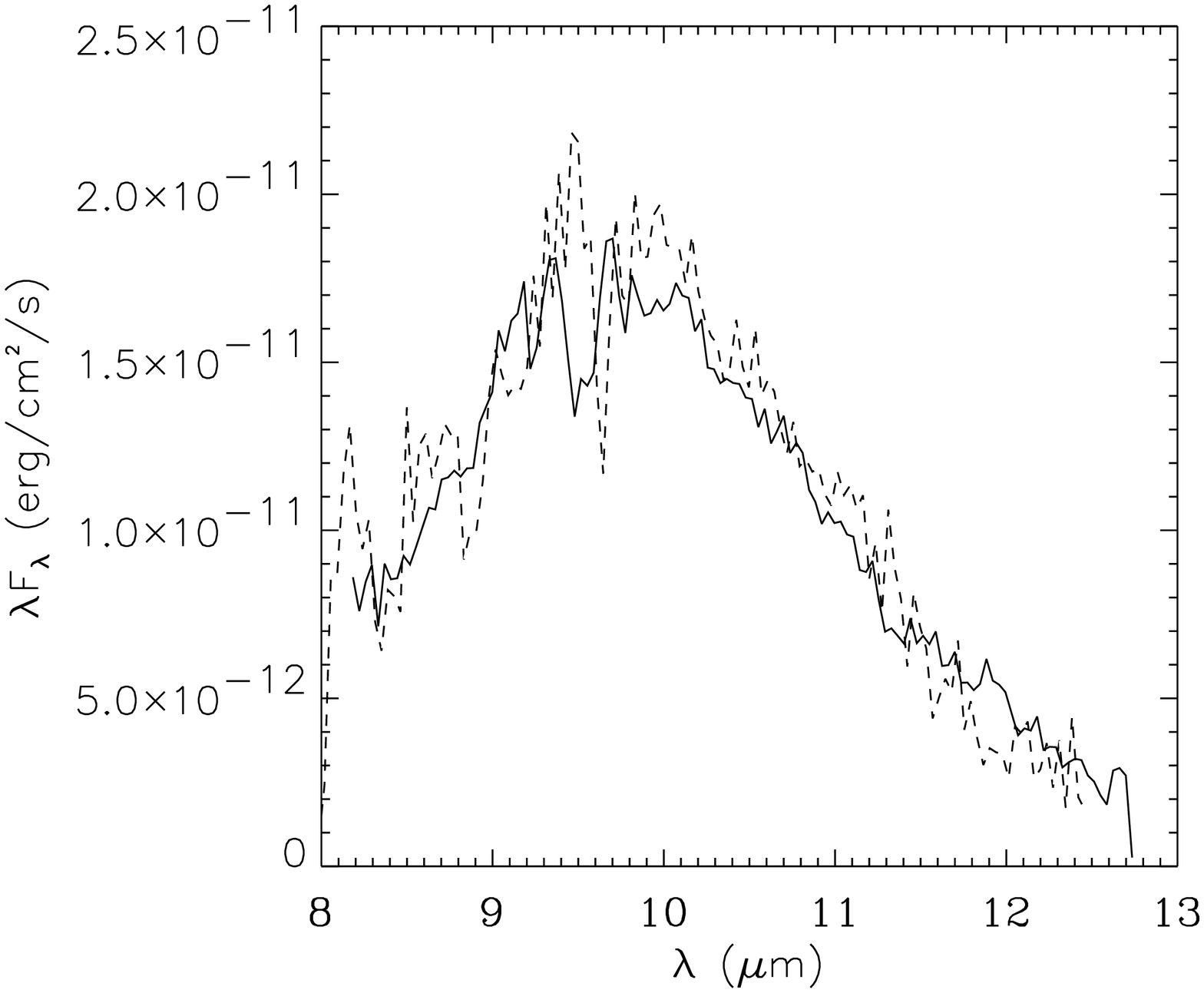}  
\vspace{0.3cm}
\figcaption{Searches for variability.  The left panel shows the silicate emission toward HD 163296 observed
on 1999 August 23 (solid line) and 2000 June 20 (dashed line).  The right panel shows the silicate emission observed
toward LkCa 15 on 1999 November 30 (solid line) and 2000 November 8 (dashed line).
The spectra from two observing runs are overlaid, showing
that there is no significant variability (above the error) in the shape of the silicate emission feature for 
these two sources on a yearly timescale. 
The continuum was not subtracted for these spectra.
\label{fig:variability}}
\end{figure*}

Long term monitoring of several sources, including DG Tau \citep{wooden_00} and  HD 45677 \citep{sitko_94}, has revealed dramatic variations in the strength and shape of silicate features over periods of months to years.  Observations of DG Tau revealed that although the dust continuum remained fairly constant, the silicate feature evolved from nearly featureless continuum to emission from crystalline and amorphous silicates to emission from primarily amorphous silicates and, finally, to absorption from amorphous silicates between 1996 and 2000. \citet{wooden_00} indicate that these variations in the silicate feature are due to dust in the optically thin disk surface, and that the continuum is produced by the underlying disk midplane \citep[e.g.,][]{calvet_92}.  For HD 45677, the flux of the silicate emission feature was observed to decrease by $\sim$30\% between 1980 and 1992, coincident with an increase in the stellar flux of $\sim$60\% \citep{sitko_94}.  The authors suggest a scenario in which long-term (decade-long) variations in UV and optical flux are due to a clear hole moving into the line of sight, which is possibly coupled with a loss of small grains, as indicated by the change in flux of the silicate feature. 

Multiple spectra were obtained toward a few sources in our sample, permitting study of the variability in shape and strength of the silicate emission.
Figure~\ref{fig:variability} shows the silicate emission toward HD 163296 and LkCa 15 for two observations taken approximately one year apart.  Spectra from the two observing runs are overlaid, showing that they are consistent (within the errors).  The double peaked structure seen in the spectrum of HD 163296 has been previously observed in spectra obtained by ISO \citep[1996 October 14]{bouwman_01} and ground-based observations \citep[1996 October 10]{sitko_99} and these spectra do not differ significantly from our observations.  

HD 179218, Hen 3-600A and MWC 480 have also been observed on more than one occasion reported in the literature.  
The previous observations of HD 179218 \citep[ISO LWS, 1996 October 5]{bouwman_01} are virtually identical to our observations (2000 February 21). The spectra depict very similar peak strengths  and shapes.  
MWC 480, however, exhibits very different emission in the \citet{sitko_99} (IRTF BAAS, 1996 October 14) study than in our spectrum (1999 November 30). In particular, the 11.3 $\mu$m feature (crystalline forsterite) is much less prominent in our observations and there is an additional peak at $\sim$10.5 $\mu$m (near the position of crystalline enstatite).  However, the (peak-to-continuum) strength of the feature is the same to within $\sim$15$\%$.   The observations of Hen 3-600A by \citet{honda_03} (Subaru COMICS, 2001 December 27) also suggest strong evidence of variability when compared to our data (Figure~\ref{fig:tts}, 2000 February 21).  Our spectrum shows much less emission at 9.2 $\mu$m and possibly 10.3 $\mu$m (both attributed to enstatite), with the peak at 11.3 $\mu$m (forsterite) dominating the spectrum.  There also appears to be an accompanying decrease in the continuum or in the strength of emission near 12.5 $\mu$m from 2000 to 2001.

There is no obvious trend in the variability with respect to spectral parameters or silicate emission feature shape from these observations.  HD 179218 and Hen 3-600A have similar emission features, ages and are in similar evolutionary stages (although their mass differs by a factor of 2).Yet the spectrum of Hen 3-600A is highly variable on 1 yr timescales while the spectrum of HD 179218 has not changed in $>$~3 yr.  Although HD 163296 and MWC 480 possessed similar spectra in 1996 \citep{sitko_99}, the spectra for these sources are quite different from each other in 1999 (our data). MWC 480 has undergone changes in the spectrum that indicate conversion between the crystalline forms of forsterite (Mg$_2$SiO$_4$) and enstatite (MgSiO$_3$).  This conversion, also seen in the Hen 3-600A spectrum, is the opposite of the expected transition with time  during the standard annealing process \citep[see][]{gail_98}.  

Studies of the variability of silicate emission may provide valuable information about the mechanisms of grain growth and/or crystallization that affect the feature shape.
However, variability of the silicate emission feature has been observed toward only a few objects, and rarely in a comprehensive program designed to monitor the accompanying variability of the star and/or SED.  Dedicated long term monitoring of the silicate feature toward a larger sample of objects is required to understand the cause of silicate feature variability in disks.

\section{SED References}

The SEDs presented in Figures~\ref{fig:ysolow}-\ref{fig:debhigh} were compiled from photometric observations in the literature.  The references for each source are listed below.

\renewcommand{\labelitemi}{}
\begin{itemize}
\item 49 Ceti:  The SED includes optical photometry from \citet{hayes_75}, UBV photometry from \citet{rybka_69} and \citet{stoy_69}, JHKLM photometry from \citet{eiroa_01} and \citet{sylvester_96}, mid-IR photometry from \citet{thi_01}, and IRAS observations from \citet{IRAS}.

\item HD 17925: The SED includes optical photometry from \citet{hayes_75}, UBVRIJHKL photometry from \citet{johnson_66}, \citet{cutispoto_01}, \citet{bessel_90} and \citet{johnson_68}, N-band photometry from \citet{metchev_04}, and IRAS observations from \citet{IRAS} as presented in \citet{habing_d01}.

\item HD 22049: The SED includes optical photometry from \citet{hayes_75}, Geneva 7-color photometry (U,B1,B,B2,V1,V,G) from \citet{rufener_76}, UBVRIJHKL photometry from \citet{johnson_66}, \citet{mendoza_70} and \citet{glass_75}, IRAS photometry from \citet{IRAS} and \citet{habing_d01}, and 800--1300 $\mu$m photometry from \citet{chini_90}.

\item IRAS 04016+2610: The SED includes optical, JHKL, (sub)millimeter photometry from \citet{kenyon_93} and references therein, 1--2 $\mu$m photometry from \citet{padgett_b99}, mid-IR photometry from \citet{myers_87} and this paper, submillimeter photometry from \citet{young_03} and \citet{hogerheijde_s00}, and millimeter photometry from \citet{moriarty-Schieven_94}, \citet{motte_andre_01}, \citet{hogerheijde_97}, \citet{ohashi_96}, \citet{saito_01} and \citet{lucas_00}.

\item IRAS 04108+2803: The SED includes optical, JHKL, (sub)millimeter photometry from \citet{kenyon_93} and references therein, mid-IR photometry from \citet{myers_87} and this paper, IRAS photometry from \citet{IRAS}, submillimeter photometry from \citet{young_03}, and millimeter photometry from \citet{moriarty-Schieven_94}, \citet{motte_andre_01} and \citet{ohashi_96}.

\item IRAS 04169+2702: The SED includes K-band photometry from \citet{tamura_91}, K-band, millimeter and submillimeter photometry from \citet{barsony_92}, IRAS photometry from \citet{IRAS}, submillimeter photometry from \citet{young_03}, millimeter photometry from \citet{moriarty-Schieven_94}, \citet{motte_andre_01}, and radio photometry from \citet{saito_01}.

\item IRAS 04181+2654 A/B:  The SEDs include JHK band photometry from \citet{kenyon_93} and references therein, IRAS photometry from \citet{IRAS}, and submillimeter photometry from \citet{moriarty-Schieven_94}.

\item IRAS 04239+2436: The SED includes JHKL band photometry from \citet{kenyon_93} and \citet{padgett_b99}, IRAS photometry from \citet{IRAS}, submillimeter photometry from \citet{young_03} and \citet{moriarty-Schieven_94}, millimeter photometry from \citet{motte_andre_01}, \citet{ohashi_96}, and \citet{saito_01}, and radio photometry from \citet{lucas_00}.

\item IRAS 04248+2612: The SED includes JHKL band photometry from \citet{kenyon_93} and \citet{padgett_b99}, N-band photometry from this paper, IRAS photometry from \citet{IRAS}, submillimeter photometry from \citet{young_03}, \citet{moriarty-Schieven_94}, and \citet{dent_m98}, millimeter photometry from \citet{motte_andre_01} and \citet{ohashi_96}, and radio photometry from \citet{lucas_00}.

\item IRAS 04264+2433: The SED includes JHKL band photometry from \citet{kenyon_93} and references therein, N-band photometry from this paper, IRAS photometry from \citet{IRAS},  submillimeter photometry from \citet{young_03}, and millimeter photometry from \citet{motte_andre_01}.

\item Haro 6-10 A/B:  The SEDs include BVRIJH and mid-IR photometry from \citet{myers_87}, KLM photometry \citet{kenyon_93} and \citet{leinert_89}, IRAS photometry from \citet{IRAS}, and (sub)millimeter photometry from \citet{chandler_b98} and \citet{kenyon_93}.

\item IRAS 04287+1801: The SED includes JHKLM band photometry from \citet{kenyon_93}, N-band photometry from this paper, IRAS photometry from \citet{IRAS}, and (sub)millimeter photometry from \citet{chandler_r00}, \citet{moriarty-Schieven_94}, \citet{motte_andre_01}, \citet{hogerheijde_97}, \citet{ohashi_96} and \citet{saito_01}.

\item IRAS 04295+2251: The SED includes JHKL and mid-IR photometry from \citet{kenyon_93} and references therein, N-band photometry from this paper, IRAS photometry from \citet{IRAS}, submillimeter photometry from \citet{young_03} and \citet{moriarty-Schieven_94}, and millimeter photometry from \citet{motte_andre_01} and \citet{ohashi_96}.

\item AA Tau: The SED includes UBV photometry from \citet{joy_49}, mid-IR photometry from \citet{thi_01} and \citet{metchev_04}, IRAS photometry from \citet{IRAS}, and (sub)millimeter photometry from \citet{dutrey_g96}, \citet{beckwith_sargent_91}, \citet{beckwith_s90}, and \citet{qi}.

\item LkCa 15: The SED includes UBVRIJHKLN photometry from \citet{kenyon_h95} and references therein, mid-IR photometry from \citet{thi_01}, IRAS photometry from \citet{IRAS}, and millimeter photometry from \citet{osterloh_b95}, \citet{duvert_g00}, and \citet{qi}.

\item IRAS 04381+2450: The SED includes JHKL and mid-IR photometry from \citet{kenyon_93} and references therein, IRAS photometry from \citet{IRAS}, and (sub)millimeter photometry from \citet{young_03}, \citet{moriarty-Schieven_94},  \citet{hogerheijde_s00}, and \citet{hogerheijde_97}.

\item IRAS 04489+3042: The SED includes JHKL and mid-IR photometry from \citet{kenyon_93} and references therein, N-band photometry from this paper, and millimeter photometry from \citet{motte_andre_01}.

\item GM Aur: The SED includes UBV photometry from \citet{bastian_79}, UBVRIJHK photometry from \citet{kenyon_h95}, mid-IR photometry from \citet{thi_01}, IRAS photometry from \citet{IRAS}, and (sub)millimeter photometry from \citet{beckwith_sargent_91}, \citet{dutrey_g96}, \citet{beckwith_s90}, and \citet{qi}.

\item MWC 480: The SED includes optical photometry from \citet{eimontas_98} and \citet{haupt_74}, HKL photometry from \citet{allen_73}, mid-IR photometry from \citet{thi_01}, IRAS photometry from \citet{IRAS}, and millimeter photometry from \citet{mannings_s97} and \citet{qi}.

\item BN : The SED includes near-IR photometry from \citet{hillenbrand_01} and \citet{dougados_93}, mid-IR photometry from \citet{gezari_98}, millimeter photometry from \citet{plambeck_95}, and radio photometry from \citet{felli_93} and \citet{menten_95}.

\item NGC 2024 IRS2: The SED includes optical photometry from Hillenbrand, unpublished, CCD, JHKL photometry from \citet{haisch_l01}, mid- and far-IR photometry from \citet{grasdalen_74}, \citet{haisch_l01}, \citet{thronson_78,thronson_84}, millimeter photometry from \citet{wilson_95} and \citet{eisner_03}, and radio photometry from \citet{gaume_92}, \citet{rodriguez_03}, and \citet{snell_86}.

\item Mon R2 IRS3:  The SED includes JHKLM photometry from \citet{preibisch_02}, \citet{carpenter_97}, and \citet{koresko_93}, mid-IR photometry from \citet{walsh_01} and \citet{persi_96}, IRAS photometry from \citet{IRAS}, (sub)millimeter photometry from \citet{jenness_95} and \citet{henning_92}, and radio photometry from \citet{tofani_95}.

\item HD 233517: The SED includes optical photometry from \citet{sylvester_96} and \citet{miroshnichenko_96}, JHKLL'MN photometry from \citet{skinner_95} and \citet{sylvester_96}, IRAS photometry from \citet{IRAS}, and millimeter photometry from \citet{duncan_90} and \citet{sylvester_01}.

\item Hen 3-600A: The SED includes BVRI photometry from \citet{gregorio-hetem_92} and \citet{torres_03}, JHKLMN photometry from \citet{geoffray_m01}, \citet{webb_99}, \citet{weintraub_00}, \citet{jayawardhana_99}, and \citet{metchev_04}, and IRAS photometry from \citet{IRAS}.

\item HD 102647: The SED includes optical photometry from \citet{hayes_75}, Geneva 7-color photometry (U,B1,B,B2,V1,V,G) from \citet{rufener_76}, UBVRIJK photometry from \citet{johnson_66}, mid-IR photometry from \citet{jayawardhana_01} and \citet{metchev_04}, IRAS photometry from \citet{IRAS}, and 60 $\mu$m ISO photometry from \citet{habing_d01}.

\item HR 4796A: The SED includes optical photometry from \citet{hayes_75},  UBVRIJHK photometry from \citet{jura_z93}, KLMN and mid-IR photometry from \citet{fajardoacosta_98}, N-band photometry from \citet{metchev_04}, IRAS photometry from \citet{IRAS}, and submillimeter photometry from \citet{jura_95}.

\item IRAS 14050-4109: The SED includes optical photometry from \citet{gregorio-hetem_02}, N-band photometry from \citet{metchev_04}, and IRAS photometry from \citet{IRAS}.

\item HD 163296: The SED includes optical photometry from \citet{jamar_95}, \citet{haupt_74}, \citet{the_85}, and \citet{malfait_98}, and \citet{oudmaijer_01}, UBVRI photometry from \citet{hillenbrand_s92}, JHKLMN  photometry from \citet{berilli_92}, \citet{eiroa_01}, and \citet{allen_73}, mid-IR photometry from \citet{thi_01} and \citet{jayawardhana_01}, IRAS photometry from \citet{IRAS}, and (sub)millimeter photometry from \citet{mannings_e94} and \citet{qi}.

\item HD 179218: The SED includes optical photometry from \citet{hayes_75}, UPXYZVS photometry from \citet{eimontas_98},  UBVRIJHKLMN photometry from \citet{miroshnichenko_m99}, \citet{malfait_98}, and \citet{lawrence_90}, IRAS photometry from \citet{IRAS}, submillimeter photometry from \citet{meeus_01}, and millimeter photometry from \citet{mannings_s00}.

\item WW Vul: The SED includes UPXYZVS photometry from \citet{eimontas_98}, UBVR photometry from \citet{herbst_s99}, JHKL photometry from \citet{eiroa_01} and \citet{glass_p74}, mid-IR photometry from \citet{thi_01},  IRAS photometry from \citet{IRAS}, and millimeter photometry from \citet{natta_g97,natta_p01}.

\item HD 184761: The SED includes optical photometry from \citet{hayes_75}, UBVRIJHK photometry from \citet{miroshnichenko_m99}, and IRAS photometry from \citet{IRAS}.

\item HD 216803: The SED includes  UPXYZVS photometry from \citet{eimontas_98}, UBVRI photometry from \citet{johnson_66} and \citet{bessel_90}, mid-IR photometry from \citet{fajardoacosta_99} and \citet{metchev_04},  IRAS photometry from \citet{IRAS}, 60 $\mu$m ISO photometry from \citet{habing_d01}, and millimeter photometry from \citet{weintraub_94}.
\end{itemize}

\end{document}